\documentclass[journal]{IEEEtran}
\usepackage{float}
\usepackage{amsmath,amsfonts}
\usepackage{algorithmic}
\usepackage{algorithm}
\usepackage{array}
\usepackage[caption=false,font=normalsize,labelfont=sf,textfont=sf]{subfig}
\usepackage{textcomp}
\usepackage{stfloats}
\usepackage{url}
\usepackage{verbatim}
\usepackage{graphicx}
\usepackage{cite}
\usepackage{enumitem} 
\usepackage{bm}
\usepackage{mathtools}
\usepackage{amssymb}
\usepackage{siunitx} 
\usepackage{booktabs}
\usepackage{multirow}
\usepackage[colorlinks=true,linkcolor=blue,citecolor=blue,urlcolor=blue]{hyperref}
\newcommand{\slsh}{\;/\;}
\begin{document}

\title{Cascaded TD3-PID Hybrid Controller for Quadrotor Trajectory Tracking in Wind Disturbance Environments}



\author{Yukang~Zhang, Shuqi~Chai, Yuhang~Zhang, Danlan~Huang, and Quanbo~Ge%
\thanks{Yukang Zhang is with the School of Logistics Engineering, Shanghai Maritime University, Shanghai 200135, China (e-mail: 202430210046@stu.shmtu.edu.cn).}%
\thanks{Shuqi Chai is with the Shenzhen Research Institute of Big Data, Shenzhen 518000, China, also with the Chinese University of Hong Kong, Shenzhen, Shenzhen 518172, China (e-mail: schai@sribd.cn).}%
\thanks{Yuhang Zhang is with the School of Information Engineering, Henan University of Science and Technology, Luoyang 471023, China (e-mail: yhZhang@stu.haust.edu.cn).}%
\thanks{Danlan Huang is with the Key Laboratory of Universal Wireless Communications, Ministry of Education, Beijing University of Posts and Telecommunications, Beijing 100876, China (e-mail: huangdl@bupt.edu.cn).}%
\thanks{Quanbo Ge is with the School of Automation, Nanjing University of Information Science and Technology, Nanjing 210044, China, also with the Shenzhen Research Institute of Big Data, Shenzhen 518000, China, and also with the Jiangsu Collaborative Innovation Center on Atmospheric Environment and Equipment Technology (CICAEET), Nanjing 210044, China (e-mail: geqb@nuist.edu.cn).}%
}

\maketitle

\begin{abstract}
This work presents a cascaded hybrid control framework for quadrotor trajectory tracking under nonlinear dynamics and external disturbances. In quadrotor systems, the altitude and attitude channels exhibit fast, structured dynamics that are well suited to reliable regulation, whereas horizontal-position control is more strongly affected by coupling effects, uncertainty, and disturbances, so that neither pure feedback control nor purely learning-based control alone is equally well suited to all channels. Accordingly, the proposed framework augments conventional proportional-integral-derivative (PID) stabilization for altitude and attitude control with an enhanced Twin Delayed Deep Deterministic Policy Gradient (TD3) agent incorporating a multi-Q-network structure, thereby improving horizontal-position control under severe disturbances. To further strengthen disturbance rejection in altitude and attitude control, a hybrid disturbance observer (HDOB) using low-pass and exponential moving average filtering is embedded in the control loops. The proposed TD3 enhancements are verified through ablation studies, and both numerical simulations and real-world flight tests on the quadrotor platform demonstrate that the proposed method achieves more accurate and robust trajectory tracking under wind disturbances than baseline approaches.
\end{abstract}


\section{Introduction}
\label{sec:I}
\IEEEPARstart{Q}{uadrotors}, as unmanned aerial vehicles (UAVs) with simple structures and high maneuverability, have demonstrated broad prospects in industrial and commercial applications. 
Compared to fixed-wing and flapping-wing UAVs, vertical takeoff and landing, stable hovering in confined spaces, and agile pose control can be achieved by quadrotors, which makes them well suited to complex environments.
With advances in sensor integration and wireless communication, quadrotors have shown remarkable performance in both traditional and emerging applications, including autonomous inspection \cite{refX}, logistics transportation \cite{ref2}, emergency rescue \cite{ref3}, and swarm flight \cite{ref4}. 
These developments have driven the need for increasingly precise and safe trajectory-tracking control methods for quadrotor systems.

Despite the widespread adoption of cascaded architectures in quadrotor trajectory-tracking control, owing to their clear hierarchical separation of control objectives and favorable dynamic decoupling \cite{ref5}, \cite{ref6}, \cite{ref7}, \cite{ref8}, traditional cascaded controllers may still face limitations under complex external disturbances. 
Their robustness can degrade in the presence of strong nonlinear aerodynamic effects, model uncertainties, and disturbance coupling, while some advanced designs remain difficult to implement in practice. Representative cascaded approaches include PID control \cite{ref9}, \cite{ref10}, \cite{ref11}, sliding-mode control (SMC) \cite{ref12}, \cite{ref13}, and model predictive control (MPC) \cite{ref14}, \cite{ref15}. 
To improve disturbance rejection and tracking performance, \cite{ref16} proposed a position-loop finite-frequency $H_{\infty}$ cascaded PD--PID controller for finite-frequency disturbances, \cite{ref17} developed a cascaded nonlinear PID controller to improve tracking accuracy under actuator disturbances, Xu et al. \cite{ref18} combined an attitude-loop active disturbance rejection controller (ADRC) with a position-loop backstepping sliding mode controller (BSMC) to handle uncertainties and external disturbances, and Wang et al. \cite{ref19} introduced UDE-based cascaded control for quadrotor field inspection tasks. Recent hierarchical designs have also considered quadrotor transport systems \cite{ref20} and safety-critical control based on ECBF-enabled cascaded architectures \cite{ref21}. 
However, robustness under strong wind disturbances remains insufficiently addressed in \cite{ref20}, while physical-platform validation is still absent in \cite{ref21}.

In contrast to conventional model-based approaches, which often struggle to maintain robustness under significant aerodynamic uncertainties, reinforcement learning (RL) offers a data-driven alternative because control policies are learned from interaction data rather than specified entirely by an accurate model. This enables RL to better accommodate modeling errors and time-varying disturbances. 
Accordingly, RL-based methods have been explored for quadrotor pose regulation and trajectory tracking. Han et al. \cite{ref22} developed a cascaded deep RL framework under the small-angle assumption, and Lin et al. \cite{ref23} improved tracking performance through experience replay and temporal-difference learning; 
however, these tracking-oriented designs mainly emphasized performance improvement rather than explicit disturbance robustness. A hybrid RL-control framework was investigated in \cite{ref24}, where fast convergence and reduced tracking errors were achieved on micro--quadrotor platforms, although validation remained limited to a specific platform. 
For more demanding scenarios, \cite{ref25} enabled high-speed flight in confined environments through curiosity-driven and branched exploration, but at the cost of onboard deployment to meet aggressive flight requirements, whereas \cite{ref26} improved tracking under uncertainties while requiring partial system knowledge and without considering actuator saturation. 
These findings suggest that pure RL and its existing variants still face difficulty in simultaneously ensuring robustness, stability, and practical applicability under disturbances. This motivates the cascaded hybrid control architecture proposed in this work.

Furthermore, owing to the strong nonlinearity and coupling of quadrotor systems, trajectory tracking remains highly sensitive to external disturbances, which has motivated various disturbance-rejection and robustness-enhancement methods. Sun et al. \cite{ref27} integrated robust signal compensation with backstepping control to address tracking challenges caused by nonlinearity, uncertainties, and external disturbances, reducing tracking errors to a neighborhood of zero under complex disturbances. 
Estimator- and observer-based strategies have also been explored. In \cite{ref28}, a cascaded control strategy incorporating an uncertainty and disturbance estimator (UDE) was proposed to handle model uncertainties, external disturbances, and Euler-angle singularities during quadrotor trajectory tracking, while \cite{ref29} developed a disturbance-rejection scheme that combined a Lipschitz unknown-input state observer with a PD controller for multirotor tracking under wind disturbances, 
although it depends strongly on model accuracy. Hua et al. \cite{ref30} further integrated a disturbance observer with a Lyapunov-based nonlinear design to mitigate external disturbances and constrain tracking errors; however, the approach involves numerous parameters that require careful tuning. Accordingly, to improve disturbance rejection while preserving the overall merits of the hybrid cascaded framework, a structurally simple hybrid disturbance observer is developed for the altitude and attitude control loops.

To address these challenges, a cascaded hybrid controller, termed CTPH, is proposed for quadrotor trajectory tracking under nonlinear dynamics and wind disturbances. The main contributions of this work are summarized as follows:
\begin{itemize}
  \item \textbf{Hybrid cascaded architecture:} A cascaded TD3-PID hybrid framework is established through functional decomposition. PID controllers are assigned to the altitude and attitude channels with relatively fast and structured dynamics, whereas the TD3 controller is used for horizontal-position control, which is more strongly affected by coupling effects, model uncertainty, and time-varying disturbances. This channel-wise design allows the complementary strengths of reliable feedback regulation and learning-based adaptability to be jointly exploited.
  \item \textbf{Enhanced TD3-based horizontal position control:} Several enhancements are introduced to improve the TD3 controller for disturbed horizontal-position regulation. Specifically, a multi-Q network alleviates overestimation bias and improves learning stability, a PID-based expert policy guides early-stage exploration, and a reward-filtered dual experience replay mechanism improves sample utilization.
  \item \textbf{Hybrid disturbance observer for PID-governed channels:} A structurally simple hybrid disturbance observer is developed for the altitude and attitude loops by combining low-pass and exponential moving average filtering with weighted fusion. This design improves disturbance estimation and rejection while preserving practical simplicity.
  \item \textbf{Comprehensive experimental validation:} The proposed CTPH framework is evaluated through ablation studies, numerical simulations, and real-world Crazyflie flight experiments. The results verify the effectiveness of the framework and demonstrate stable and accurate trajectory tracking under external disturbances.
\end{itemize}

The remainder of this article is organized as follows. Section~\ref{sec:II} introduces the preliminaries and problem formulation. Section~\ref{sec:III} presents the horizontal-position TD3 controller together with its associated enhancements, 
while Section~\ref{sec:IV} presents the PID controllers and hybrid disturbance observer. Section~\ref{sec:V} provides implementation details and reports both simulation and real-world flight results. Finally, Section~\ref{sec:VI} concludes this article.

\section{Preliminaries and Problem Formulation}
\label{sec:II}
This section first introduces the dynamics of the quadrotor UAV in Section~\ref{sec:II-A} and then formally formulates the control problem in Section~\ref{sec:II-B}, thereby clarifying the control objective for the subsequent controller design. 
Finally, the motivation and rationale behind the layered control architecture of the quadrotor system are presented in Section~\ref{sec:II-C}.

\subsection{Mathematical Model of the Quadrotor}
\label{sec:II-A}
\begin{figure}[!t]
\centering
\includegraphics[width=2.5in]{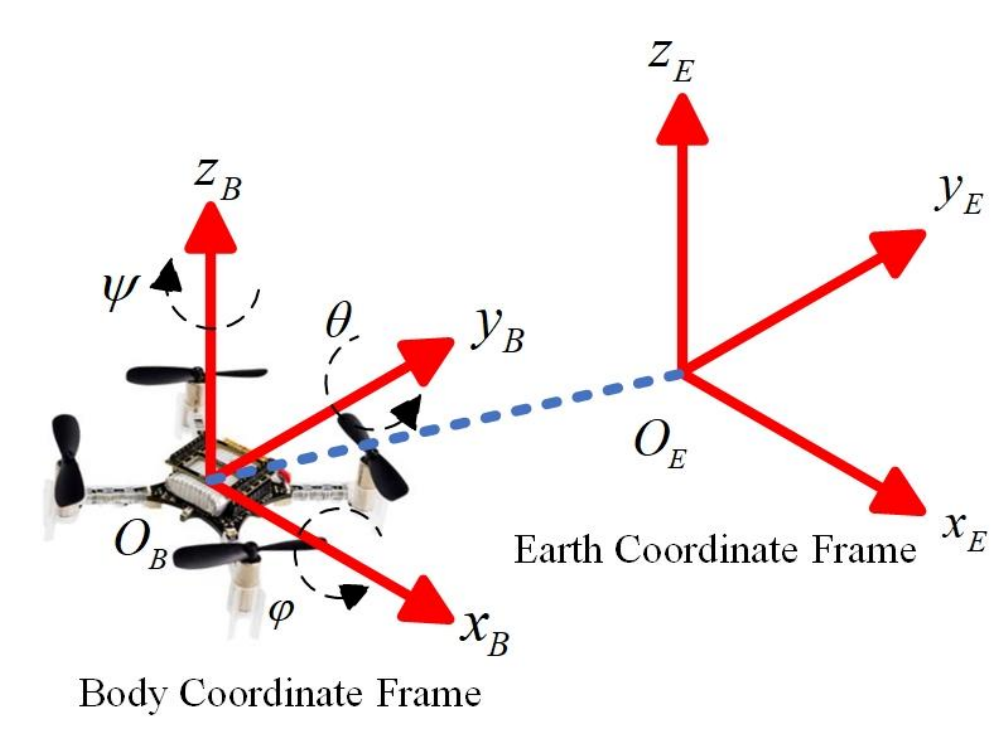}
\caption{Coordinate frame of the quadrotor. Earth coordinate frame and the body fixed frame.}
\label{fig:quadrotor}
\end{figure}

The quadrotor is a typical underactuated system with six degrees of freedom: three positional ($x, y, z$) and three attitude ($\varphi, \theta, \psi$), but only four independent control inputs. Its dynamics result from complex coupling among aerodynamic forces, mechanical structures, and control systems.
The mathematical modeling of quadrotor aircraft is usually defined in two coordinate systems, which are the Earth coordinate frame ($O_{E}x_{E}y_{E}z_{E}$) and the body coordinate frame ($O_{B}x_{B}y_{B}z_{B}$), as shown in Fig.~\ref{fig:quadrotor}.

For quadrotors with an X-configuration, the dynamic model can be formulated based on \cite{ref31} as:
\begin{equation}
\begin{aligned}
    \dot{\bm{\eta}} &= \bm{v} \,, \\
    m \dot{\bm{v}} &= \bm{R}_b^e \bm{F} + m \bm{g} \,, \\
    \dot{\bm{R}}_b^e &= \bm{R}_b^e \, \bm{\hat{\omega}}^b \,, \\
    \bm{I} \dot{\bm{\omega}} &= -\bm{\omega} \times (\bm{I} \bm{\omega}) + \bm{M} \,, \\
    \dot{\bm{\zeta}} &= \bm{\Lambda}\bm{\omega} \,,
\end{aligned}
\label{eq:dynamics}
\end{equation}
where $\bm{\eta} = [x, y, z]^{\mathrm{T}}$ and $\bm{\zeta} = [\varphi, \theta, \psi]^{\mathrm{T}}$ are the position and attitude of the quadrotor,
$m$ denotes the quadrotor mass and $\bm{g}=[0, 0, -g]^{\mathrm{T}}$ is the vector of gravity acceleration. $\bm{v}=[v_{x}, v_{y}, v_{z}]^{\mathrm{T}}$ is the linear velocity
and $\bm{\omega} = [\omega_x,\omega_y,\omega_z]^{\mathrm T}$ denotes the angular velocity of the quadrotor expressed in the body frame.
$\bm{F}=[0, 0, f]^{\mathrm{T}}$ and $\bm{M}=[M_{x}, M_{y}, M_{z}]^{\mathrm{T}}$ are defined as force and torque applied to the quadrotor UAV.
The moment of inertia matrix $\bm{I} = \mathrm{diag}(I_x, I_y, I_z)$ represents the inertia distribution in the body coordinate frame.
${\bm{R}}_b^e \in \mathrm{SO}(3)$ is the rotation matrix that describes the transformation of orientation from the body coordinate frame to the Earth coordinate frame, and it is defined as:
\begin{equation}
\bm{R}_b^e =
\scalebox{0.9}{$
\begin{bmatrix}
C\psi C\theta & C\psi S\theta S\varphi - S\psi C\varphi & C\psi S\theta C\varphi + S\psi S\varphi \\
S\psi C\theta & S\psi S\theta S\varphi + C\psi C\varphi & S\psi S\theta C\varphi - C\psi S\varphi \\
-S\theta & C\theta S\varphi & C\theta C\varphi
\end{bmatrix},$}
\label{eq:rotation_matrix}
\end{equation}
where $S(\cdot), C(\cdot)$ denote $\sin(\cdot), \cos(\cdot)$. The Euler-rate transformation matrix $\bm{\Lambda}$, which relates the Euler-angle rates to the body angular velocity, is defined as:
\begin{equation}
\bm{\Lambda} = \begin{bmatrix} 
1 & \sin \varphi \tan \theta & \cos \varphi \tan \theta \\ 
0 & \cos \varphi & -\sin \varphi \\ 
0 & \sin \varphi / \cos \theta & \cos \varphi / \cos \theta 
\end{bmatrix}.
\label{eq:Lambda}
\end{equation}

The skew-symmetric matrix $\bm{\hat{\omega}}^b \in \mathfrak{so}(3)$, obtained via the hat operator applied to the angular velocity vector $\bm{\omega}^b \in \mathbb{R}^3$, is defined as:
\begin{equation}
\bm{\hat{\omega}}^b =
\begin{bmatrix}
0 & -\omega_z & \omega_y \\
\omega_z & 0 & -\omega_x \\
-\omega_y & \omega_x & 0
\end{bmatrix}.
\label{eq:hat_matrix}
\end{equation}

Based on the above analysis, the force-moment allocation is expressed as:
\begin{equation}
\scalebox{0.9}{$
\begin{bmatrix}
f \\
M_x \\
M_y \\
M_z
\end{bmatrix}
=
\begin{bmatrix}
C_f & C_f & C_f & C_f \\
-\frac{l}{\sqrt{2}}C_f & -\frac{l}{\sqrt{2}}C_f & \frac{l}{\sqrt{2}}C_f & \frac{l}{\sqrt{2}}C_f \\
-\frac{l}{\sqrt{2}}C_f & \frac{l}{\sqrt{2}}C_f & \frac{l}{\sqrt{2}}C_f & -\frac{l}{\sqrt{2}}C_f \\
-C_M & C_M & -C_M & C_M
\end{bmatrix}
\begin{bmatrix}
\Omega_1^2 \\
\Omega_2^2 \\
\Omega_3^2 \\
\Omega_4^2
\end{bmatrix},$}
\label{eq:allocation}
\end{equation}
where $C_f$ and $C_M$ are the thrust and torque coefficients, $l$ is the distance from the rotor to the center of mass, and $\Omega_i(i=1,2,3,4)$ denotes the angular speed of rotor $i$.

\subsection{Problem Formulation}
\label{sec:II-B}

The quadrotor control problem is considered under external disturbances. The system dynamics are written as
\begin{equation}
\dot{s}(t) = \mathcal{F}(s(t), u(t)) + d(t),
\label{eq:system_disturbed}
\end{equation}
where \(s(t)\) denotes the system state, \(u(t)\) denotes the control input, \(d(t)\) denotes the external disturbance and \(\mathcal{F}(\cdot)\) represents the nominal system dynamics. The objective is to design a control law that minimizes the deviation of the quadrotor state from the desired trajectory \(s^d(t)\). This objective is expressed as
\begin{equation}
\min_{u(t)} \; J[u(t)] = \int_0^T \ell\bigl( \| s(t) - s^d(t) \| \bigr) \, dt,
\label{eq:system_u}
\end{equation}
where \(T\) is the total flight duration and \(\ell(\cdot)\) is a nonnegative error penalty function.

\subsection{Cascade Hybrid Control Scheme}
\label{sec:II-C}
\begin{figure*}[!t]   
\centering
\includegraphics[width=0.75\textwidth]{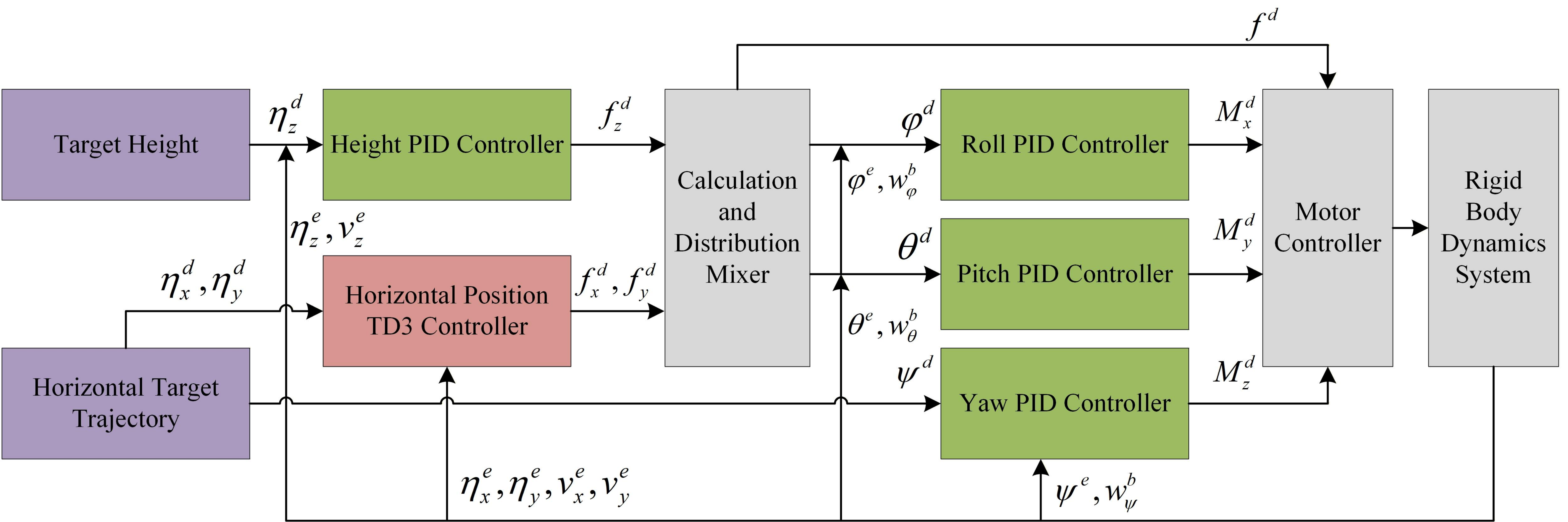}  
\caption{The proposed hybrid cascade control architecture uses the same type of controller represented by the same color regions.
A quadrotor is an underactuated system with six degrees of freedom but only four independent control inputs: total thrust and three rotational torques. It can directly track the position $x$, $y$, $z$ and yaw angle $\psi$, 
while the roll $\varphi$ and pitch $\theta$ angles are indirectly determined to support position tracking.}
\label{fig:framework}
\end{figure*}

Given the distinct dynamics of each flight channel, a control architecture based on functional decomposition is adopted, in which different controllers are assigned to different channels. 
In quadrotor systems, the altitude and attitude channels exhibit relatively fast and structured dynamics and provide the stabilization basis for position control, whereas the horizontal-position channel is more strongly affected by multivariable coupling, model uncertainty, and time-varying disturbances, making conventional feedback alone less effective. 
A hybrid control strategy integrating deep reinforcement learning with PID control is therefore adopted. Specifically, PID controllers are used for the altitude and attitude loops to ensure rapid and reliable regulation, while a TD3-based controller is introduced for horizontal-position control to improve adaptation under complex dynamic conditions. 
The overall cascade architecture of the proposed control system is illustrated in Fig.~\ref{fig:framework}.

\section{Design of Horizontal Position TD3 Controller}
\label{sec:III}
\begin{figure*}[!t] 
\centering
\includegraphics[width=0.75\textwidth]{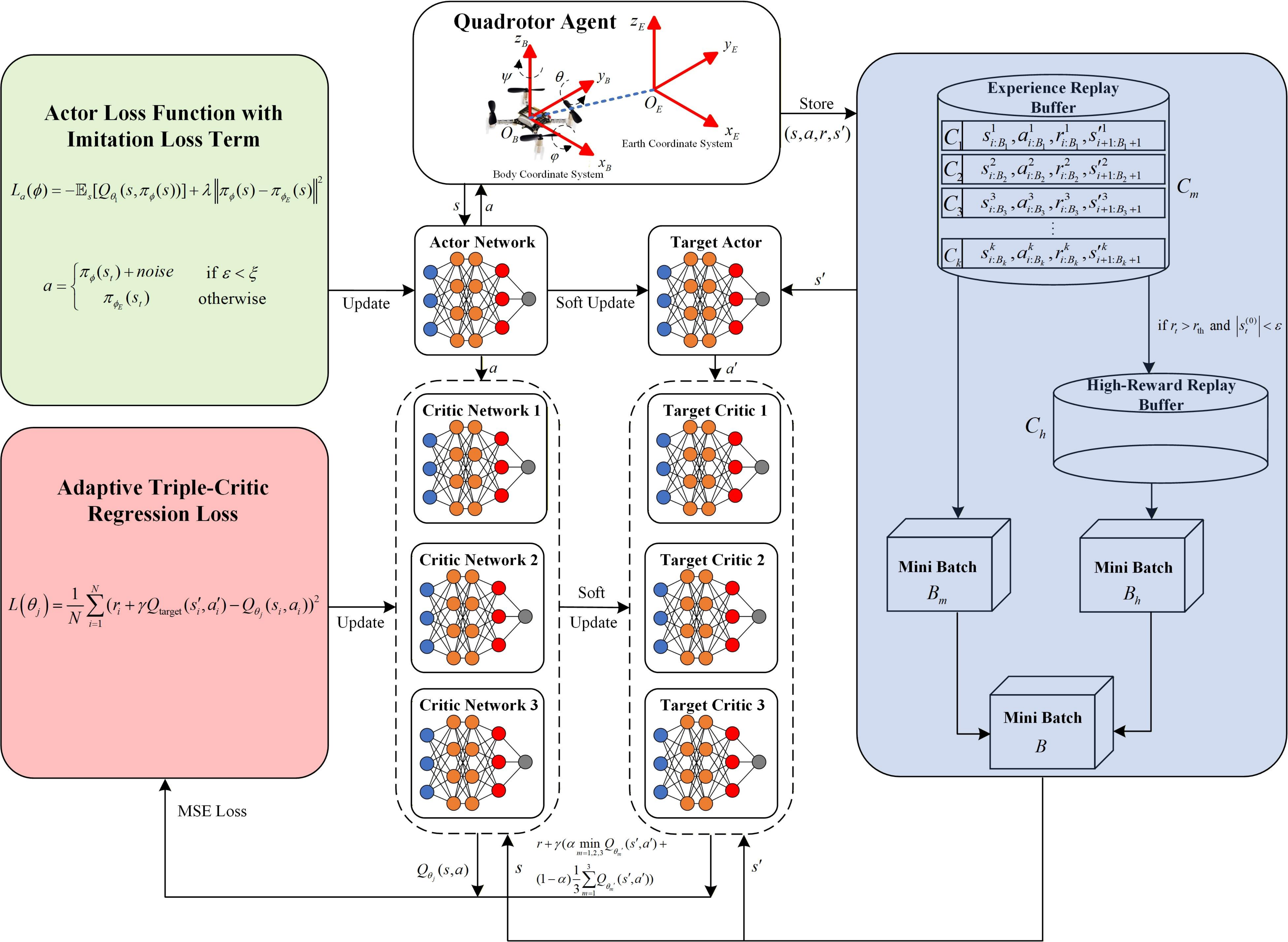}
\caption{Algorithm design of the horizontal position TD3 controller. On the basis of the original TD3 algorithm, improvements were made to action selection, actor network loss function design, target Q-value calculation, and experience replay buffer.}
\label{fig:Horizontal DRL controller}
\end{figure*}

In this section, an enhanced TD3 algorithm tailored for quadrotor horizontal position control is proposed.
Horizontal position control is more strongly affected by multivariable coupling, nonlinear aerodynamic effects, and time-varying disturbances than altitude and attitude regulation. In contrast to conventional methods that rely heavily on accurate mathematical models, a TD3-based controller can learn the horizontal control policy from interaction data and adapt to complex dynamic conditions. 
Motivated by this, an enhanced TD3 controller is developed for the horizontal position loop, with three components introduced to improve learning reliability and efficiency: an aggregated Q-network architecture, a PID-based expert policy, and reward-filtered dual experience replay. The overall algorithmic flow of the horizontal position controller is shown in Fig.~\ref{fig:Horizontal DRL controller}.

\subsection{Horizontal Translation Control Law}
\label{sec:III-A}
Since the quadrotor's thrust is constrained to the body-fixed vertical axis, horizontal acceleration is induced by adjusting roll and pitch angles to reorient the thrust vector. 
In this design, for each horizontal axis, the TD3 controller outputs a normalized action $a_i \in [-1,1]$, where $i \in \{x,y\}$, 
and this action is interpreted as the desired lateral acceleration, which is then mapped to the desired virtual horizontal force components $f_x$ and $f_y$. These virtual horizontal force components are not directly applied to the system but are used to back-calculate the desired roll angle $ \varphi^d $ and pitch angle $ \theta^d $. 
Next, the height PID controller inputs the current height $ z $ and the desired height $ z^d $. Using the desired vertical acceleration output by the height PID controller, the nominal vertical thrust command is computed as
\begin{equation}
f_z = m \cdot (g + a_z),
\label{eq:f_z}
\end{equation}
where $ g $ is the gravitational acceleration and $ a_z $ is the acceleration from the height PID controller. Subsequently, according to the equilibrium conditions, the desired roll angle $ \varphi^d $ and pitch angle $ \theta^d $ are calculated as follows:
\begin{equation}\label{eq:attitude_cmd}
\begin{aligned}
\theta^d &= \arctan\!\left(
\frac{\cos\psi^d \cdot f_x + \sin\psi^d \cdot f_y}{f_z}
\right), \\
\varphi^d &= \arctan\!\left(
\cos\theta^d \cdot
\frac{\sin\psi^d \cdot f_x - \cos\psi^d \cdot f_y}{f_z}
\right).
\end{aligned}
\end{equation}
where $\psi^{d}$ is the desired yaw angle. 

To ensure that the vertical component of the total thrust meets the flight requirements when the attitude angle deviates, it is necessary to compensate for the total thrust. Its expression is as follows:
\begin{equation}
f = \frac{m \cdot (g + a_z)}{\cos\varphi \cdot \cos\theta},
\label{eq:f}
\end{equation}
where $\varphi$ and $\theta$ are the current roll angle and pitch angle. The desired roll angle $\varphi^d$ and pitch angle $\theta^d$ are sent to the attitude PID controller. Then, the attitude PID controller generates the required desired attitude control torques $ M_x^d $ and $ M_y^d $ according to the current attitude angle $ \zeta $, 
the desired attitude angle $\zeta^d$, and the current angular velocity to adjust the fuselage angle. Finally, the total thrust $f$ obtained from \hyperref[eq:f]{(\ref*{eq:f})} and the torque $ M $ are jointly used as control inputs to the quadrotor dynamics model to drive the aircraft to generate actual propulsion and rotation responses.

\subsection{Design of State and Reward Function}
\label{sec:III-B}
The horizontal position control task is modeled as a Markov Decision Process (MDP), defined by the tuple $(\mathcal{S}, \mathcal{A}, \mathcal{P}, \mathcal{R}, \gamma, N)$, where $\mathcal{S}$ and $\mathcal{A}$ denote the state and action spaces, $\mathcal{P}(s'|s, a)$ is the transition probability, $\mathcal{R}$ is the reward function, 
$\gamma \in [0,1)$ is the discount factor, and $N$ is the time horizon. In deep reinforcement learning, neural networks are employed to approximate the policy and value functions, enabling efficient learning in high-dimensional, continuous control environments.
For the horizontal controller, the state variables for the two axes are defined as:
\begin{equation}
S_x = \{ e_x, \bar{e}_x, v_x, \theta, \dot{\theta}, a_z, M_y\},
\label{eq:S_x}
\end{equation}
\begin{equation}
S_y = \{ e_y, \bar{e}_y, v_y, \varphi, \dot{\varphi}, a_z, M_x\},
\label{eq:S_y}
\end{equation}
where \( e_x = x_d - x \) and \( e_y = y_d - y \) are the position tracking errors along the \( x \) and \( y \) directions, respectively; \( \bar{e}_x \) and \( \bar{e}_y \) represent the errors from the previous timestep, which provide temporal information to the controller; \( v_x \) and \( v_y \) are the horizontal velocities; 
\( \theta \) and \( \varphi \) denote the pitch and roll angles, and \( \dot{\theta}, \dot{\varphi} \) are their corresponding angular velocities. The vertical acceleration \( a_z \) is included to represent the desired vertical acceleration output by the altitude PID controller, while \( M_x \) and \( M_y \) represent the 
roll and pitch torques generated by the attitude PID controller.
The agent's actions correspond to the desired planar accelerations \(a_x^d\) and \(a_y^d\) along the \(x\)- and \(y\)-axes, respectively, which determine the horizontal translational motion of the quadrotor. Accordingly, the action variables are denoted by \(a_x^d\) and \(a_y^d\).

To encourage fast, stable, and accurate lateral tracking, a hybrid reward function combining dense and sparse components is proposed. Since the TD3 controller primarily aims to minimize the horizontal position error $e_i (i \in \{x, y\})$, the reward is defined as follows:
\begin{equation}
r_D = \beta_1 \cdot \exp\left(-\frac{{e_i}^2}{2}\right) - \beta_2 \cdot |e_i|,
\label{eq:r_D}
\end{equation}
which provides continuous guidance by penalizing large errors and promoting precise convergence near the target.  
A sparse bonus is added when the position error is sufficiently small:
\begin{equation}
r_S =
\begin{cases}
\beta_3, & \text{if } |e_i| < 0.01,\\
0, & \text{otherwise},
\end{cases}
\label{eq:r_S}
\end{equation}
where $\beta_1, \beta_2, \beta_3 > 0$ are scalar weights.

The final reward is computed as \( R = r_D + r_S \), following \hyperref[eq:r_D]{(\ref*{eq:r_D})} and \hyperref[eq:r_S]{(\ref*{eq:r_S})}. This hybrid mechanism combines the training stability of dense feedback with the precision guidance of sparse incentives, leading to improved convergence speed and control accuracy in practice.

\subsection{Aggregated Q-Network Architecture}
\label{sec:III-C}
Although TD3 alleviates the Q-value overestimation problem in DDPG through twin critics and minimum-based target estimation \cite{ref32}, \cite{ref33}, its dual-critic design may still be limited in complex environments by correlated estimation errors and overly conservative value approximation.

To further enhance the robustness and generalization of Q-value estimation, this work extends TD3 by incorporating an aggregated Q-network architecture. A third independently parameterized critic $Q_{\theta_3}(s, a)$ is added, and a weighted fusion mechanism is designed to combine the three critics into a more reliable target value $Q_{\text{target}}(s, a)$:
\begin{equation}
\begin{split}
Q_{\text{target}}(s, a) = \alpha \cdot \min_{i = 1, 2, 3} Q_{\theta_i'}(s, a) + (1 - \alpha) \cdot \frac{1}{3} \sum_{i = 1}^3 Q_{\theta_i'}(s, a),
\end{split}
\label{eq:Q_target}
\end{equation}
where $\alpha$ is a dynamic fusion coefficient balancing conservative and optimistic Q-value aggregation. To adaptively adjust this trade-off during training, an uncertainty-aware fusion mechanism is employed, modulating $\alpha$ based on the variance among the three critics' outputs. 
Specifically, the standard deviation of the Q-values, denoted as $\sigma_Q = \mathbb{E}_{(s, a)\sim\mathcal{D}}[\sigma(Q_{\theta_1'}(s, a), Q_{\theta_2'}(s, a), Q_{\theta_3'}(s, a))]$, is used as a proxy for uncertainty. The adaptive coefficient is defined as:
\begin{equation} 
\alpha = \alpha_{\min} + (\alpha_{\max} - \alpha_{\min}) \cdot \mu(\sigma_Q - \sigma_{\mathrm{th}}),
\label{eq:alpha}
\end{equation}
where $\mu(\cdot)$ denotes the Sigmoid function, $\sigma_{\mathrm{th}}$ is a predefined uncertainty threshold, and $0 \leq \alpha_{\min} < \alpha_{\max} \leq 1$ define the allowable range of $\alpha$. This mechanism, governed by \hyperref[eq:alpha]{(\ref*{eq:alpha})}, renders the algorithm more conservative under high uncertainty and more optimistic when value estimates are stable, thus enhancing learning stability and adaptability in complex environments.
The critic networks are then updated by minimizing the temporal-difference error $\delta_i = r_t + \gamma Q_{\text{target}}(s_{t+1}, \pi_{\phi'}(s_{t+1}) + \epsilon_s) - Q_{\theta_i}(s_t,a_t)$ for $i \in \{1,2,3\}$, where $\epsilon_s$ denotes clipped target-policy smoothing noise.

\subsection{PID-Based Expert Policy}
\label{sec:III-D}
To provide a reliable supervisory signal and ensure safe policy initialization, a PID expert controller was designed using prior domain knowledge. The proportional ($K_p^E$), integral ($K_i^E$), and derivative ($K_d^E$) gains were manually tuned based on empirical insights. 
In the quadrotor simulation environment, the PID controller was used to control the UAV, and the corresponding state-action pairs $(s_t^E, a_t^E)$ for $t=1, \dots, N$ were recorded to form the expert demonstration dataset $\mathcal{D}^E = \{(s_t^E, a_t^E)\}_{t=1}^N$. 
An MLP-based expert policy network $\pi_{\phi_E}(s)$ was then trained to mimic the PID controller by taking $s_t^E$ as input and outputting the corresponding action. The output layer employed a Tanh activation function to bound the action range. The network was optimized by minimizing the mean squared error:
\begin{equation}
L_E(\phi_E) = \frac{1}{N} \sum_{t=1}^N \left\lVert \pi_{\phi_E}(s_t^E) - a_t^E \right\rVert^2.
\label{eq:L_E}
\end{equation}

During policy execution, the action $a_t$ is selected based on the current state $s_t$ and the outputs of both the expert and TD3 policies. With probability $\xi$, the agent executes the TD3 policy $\pi_\phi(s_t)$ with added exploration noise; otherwise, it follows the fixed expert policy $\pi_{\phi_E}(s_t)$:
\begin{equation}
a_t =
\begin{cases}
\pi_\phi(s_t) + \text{noise}, & \text{if } \varepsilon < \xi\\
\pi_{\phi_E}(s_t), & \text{otherwise}
\end{cases},
\label{eq:a_t}
\end{equation}
where $\varepsilon$ is a uniformly distributed random number in the interval $[0,1)$ and $\xi$ denotes the probability of selecting the TD3 policy. Initially, $\xi = 0$, ensuring safe exploration by relying entirely on the expert policy. During training, $\xi$ is gradually increased after each step until control is fully taken over by the TD3 policy.

During the policy training phase, the expert action serves as a supervised signal incorporated into the Actor loss through a weighted imitation term:
\begin{equation}
L_{a}(\phi) = -\mathbb{E}_{s} \left[ Q_{\theta_1}(s, \pi_\phi(s)) \right] + \lambda \cdot \underbrace{\| \pi_\phi(s) - \pi_{\phi_E}(s) \|^2}_{\text{imitation loss term}},
\label{eq:L_a}
\end{equation}
where $\pi_\phi(s)$ denotes the action output by the current Actor network, $Q_{\theta_1}(s,\pi_\phi(s))$ is the first critic used for actor optimization, and $\lambda = \lambda_{T} \cdot \lambda_{P}$ is the imitation weight. The weight factor $\lambda$ is determined by a time-dependent decay factor $\lambda_{T}$ and a performance-based modulation factor $\lambda_{P}$:
\begin{equation}
\lambda_{T} = 
\begin{cases} 
{\beta}^{\frac{t}{\kappa}}, & \text{if } t < t_0 \\ 
0, & \text{otherwise} 
\end{cases}
\label{eq:lambda_T}
\end{equation}
\begin{equation}
\lambda_{P}=\frac{1}{1 + e^{-k \cdot \Delta Q}},
\label{eq:lambda_P}
\end{equation}
where $\beta \in (0, 1)$ is the decay rate, $\kappa$ is a temporal scaling factor controlling the decay speed, $\Delta Q = \mathbb{E}_{s}\!\left[ Q_{\theta_1}(s,\pi_{\phi_E}(s)) - Q_{\theta_1}(s,\pi_\phi(s)) \right]$ denotes the average Q-value difference between the expert and TD3 policies over sampled states, and $k$ shapes the steepness of the sigmoid function.  
This design, through the adaptive weight $\lambda$ in \hyperref[eq:L_a]{(\ref*{eq:L_a})}, allows the imitation loss to gradually decrease over time and adjust dynamically based on policy performance, enabling a smooth transition from expert-guided learning to autonomous optimization.

\subsection{Reward-Filtered Dual Experience Replay}
\label{sec:III-E}
To improve sample efficiency and training stability, we propose a reward-filtered dual experience replay mechanism. Unlike traditional single-buffer schemes, the proposed method maintains two separate buffers: a main replay buffer for storing all interaction data, 
and a high-reward buffer for selectively storing valuable experiences, enabling focused learning from critical transitions while preserving exploration diversity.

Let $\mathcal{B}_\mathrm{m}$ denote the main replay buffer with capacity $C_\mathrm{m}$, which stores all transitions $(s, a, r, s')$ collected during agent-environment interactions. 
In parallel, a high-reward buffer $\mathcal{B}_\mathrm{h}$ with capacity $C_\mathrm{h}$ is maintained to store transitions with high learning potential.
To determine which transitions qualify as high-reward, a sliding reward window $\mathcal{W}_r$ of length $L$ records recent instantaneous rewards. When $|\mathcal{W}_r|$ reaches the threshold $L_m$, the reward percentile threshold $r_{\mathrm{th}}$ is defined as:
\begin{equation}
r_{\mathrm{th}} = \mathrm{P}(\mathcal{W}_r, \rho),
\label{eq:r_th}
\end{equation}
where $\mathrm{P}(\cdot, \rho)$ denotes the $\rho$-th percentile of the reward distribution. A transition is considered valuable and added to $\mathcal{B}_\mathrm{h}$ if it satisfies $r_t > r_{\mathrm{th}}$ and $|s_t^{(0)}| < \epsilon$, where $s_t^{(0)}$ denotes the first dimension of the state and $\epsilon$ is a predefined state threshold.

During training, if $|\mathcal{B}_\mathrm{h}| \geq 2 N_\mathrm{h}$, a mixed mini-batch of size $B$ is formed by sampling $N_\mathrm{h}$ transitions from the high-reward buffer and $N_\mathrm{m} = B - N_\mathrm{h}$ from the main buffer. Otherwise, the entire batch is sampled from $\mathcal{B}_\mathrm{m}$.
To ensure early accumulation of high-reward experiences when the agent is still underperforming, the threshold is initialized as $r_{\mathrm{th}} = -\infty$ when $|\mathcal{W}_r| < L_m$, allowing all experiences to be stored in $\mathcal{B}_\mathrm{h}$. As training progresses, the threshold gradually increases, shifting emphasis from exploration to exploitation of high-quality behaviors.
This mechanism facilitates efficient policy refinement by increasing the frequency of valuable sample reuse while preserving exploration diversity during training.

\begin{algorithm}[!t]
\caption{Enhanced TD3 for Quadrotor Horizontal Control}
\label{alg:enhanced_td3}
\begin{algorithmic}[1]
\STATE Initialize actor $\pi_\phi(s)$ and critics $Q_{\theta_1}, Q_{\theta_2}, Q_{\theta_3}$; set target networks $\pi_{\phi'}(s)$ and $Q_{\theta_i'}$ with $\phi' \leftarrow \phi$, $\theta_i' \leftarrow \theta_i$
\STATE Initialize replay buffers $\mathcal{B}_\mathrm{m}$ and $\mathcal{B}_\mathrm{h}$
\STATE Initialize expert policy $\pi_{\phi_E}(s)$ and parameters $\xi \leftarrow 0$, $\lambda$
\FOR{episode $= 1$ to $M$}
    \STATE Observe initial state $s_0$
    \FOR{$t = 0$ to $T$}
        \STATE Select action $a_t$ by \hyperref[eq:a_t]{(\ref*{eq:a_t})} with probability $\xi$
        \STATE Compute desired attitude commands $\theta^d$, $\varphi^d$ and thrust $f$ via \hyperref[eq:attitude_cmd]{(\ref*{eq:attitude_cmd})} and \hyperref[eq:f]{(\ref*{eq:f})}
        \STATE Apply control input, observe next state $s_{t+1}$ and reward $r_t$
        \STATE Store $(s_t, a_t, r_t, s_{t+1})$ in $\mathcal{B}_\mathrm{m}$
        \IF{$r_t > r_{\mathrm{th}}$ \AND $|s_t^{(0)}| < \epsilon$}
            \STATE Store transition in $\mathcal{B}_\mathrm{h}$
        \ENDIF
        \STATE Update $r_{\mathrm{th}}$ using \hyperref[eq:r_th]{(\ref*{eq:r_th})} and the initialization rule
        \STATE Sample a mixed mini-batch from $\mathcal{B}_\mathrm{h}$ and $\mathcal{B}_\mathrm{m}$ according to the reward-filtering rule
        \STATE Compute aggregated $Q_{\text{target}}$ via \hyperref[eq:Q_target]{(\ref*{eq:Q_target})} and \hyperref[eq:alpha]{(\ref*{eq:alpha})}
        \STATE Update critics by minimizing TD error using aggregated Q-target
        \IF{$t \bmod d = 0$}
            \STATE Update actor via \hyperref[eq:L_a]{(\ref*{eq:L_a})}, with imitation term weighted by dynamic $\lambda$
            \STATE Soft-update target networks: $\phi' \leftarrow \tau \phi + (1-\tau)\phi'$, $\theta_i' \leftarrow \tau \theta_i + (1-\tau)\theta_i'$
        \ENDIF
        \STATE Update imitation weight $\lambda$ via \hyperref[eq:lambda_T]{(\ref*{eq:lambda_T})}, \hyperref[eq:lambda_P]{(\ref*{eq:lambda_P})}, and increase $\xi$ according to the exploration schedule
    \ENDFOR
\ENDFOR
\end{algorithmic}
\end{algorithm}

\subsection{Overall Training Process}
\label{sec:III-F}
The complete training procedure of the enhanced TD3 algorithm is summarized in Algorithm.~\ref{alg:enhanced_td3}, which integrates the three key components introduced in Sections~\ref{sec:III-C}-\ref{sec:III-E}. 
At each time step, the current state is fed to the actor, the selected action is converted to control commands and executed in the quadrotor dynamics, and the resulting transition is stored in the dual replay buffers. 
Mini-batches are then sampled according to the reward-filtering rule to compute the aggregated Q-target and update the critic and actor networks, after which the target networks and imitation weight are updated until convergence or the maximum number of episodes is reached.

\begin{figure*}[t!]
\centering
\includegraphics[width=0.7\textwidth]{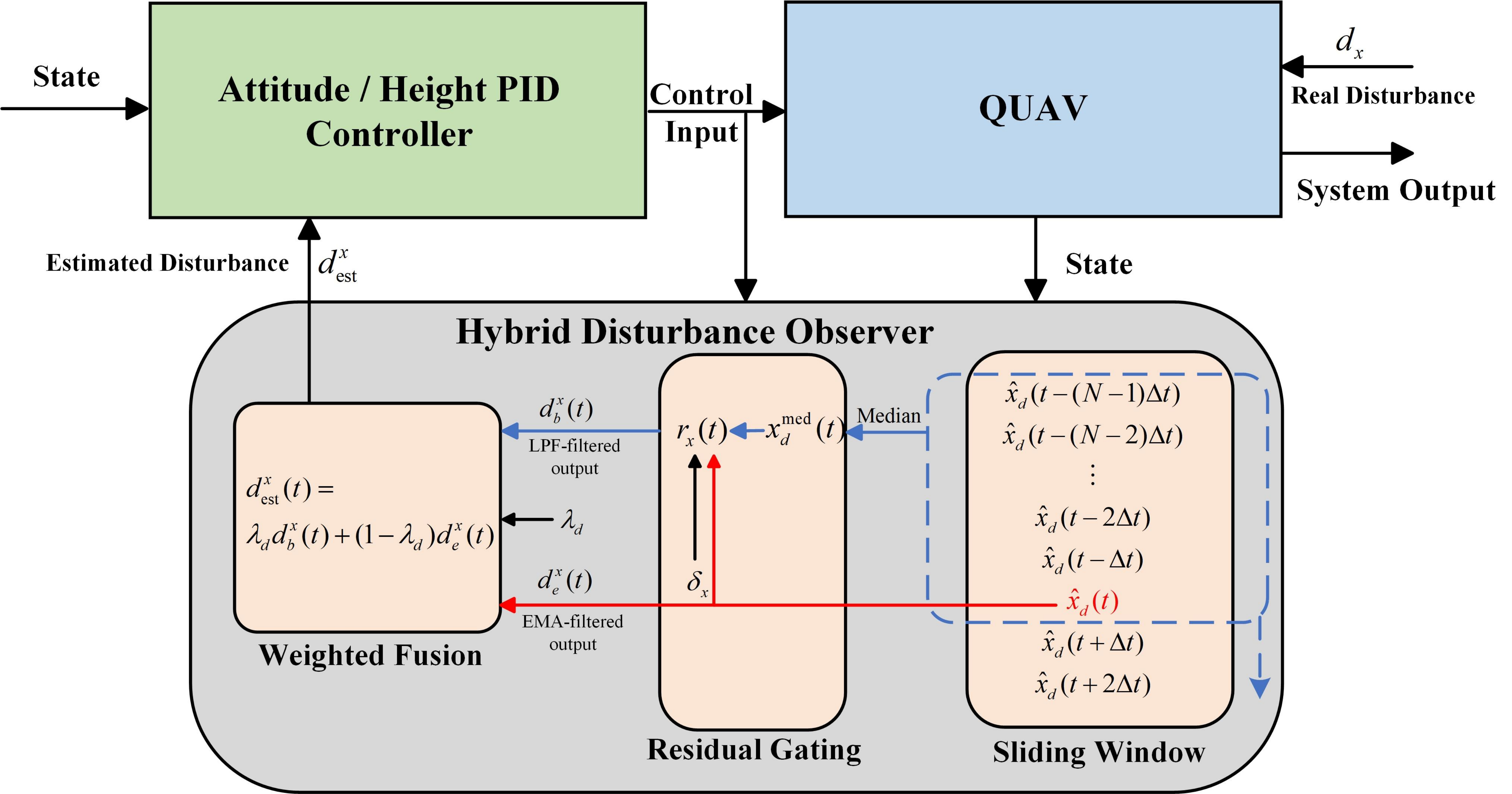}
\caption{The proposed HDOB structure combining a median filter, EMA, and IIR low-pass filter.}
\label{fig:HDOB}
\end{figure*}

\section{Design of PID Controllers and Disturbance Observers}
\label{sec:IV}
In contrast to the TD3-based horizontal position loop, the altitude and attitude loops require fast, structured, and reliable inner-loop regulation, for which PID control remains appropriate. However, these loops are still sensitive to external disturbances, which may degrade tracking performance and closed-loop stability. Motivated by this, PID controllers are adopted for altitude and attitude regulation, and a hybrid disturbance observer is introduced to improve disturbance compensation and robustness. The design of the PID controllers is presented in Section~\ref{sec:IV-A}, followed by the hybrid disturbance observer in Section~\ref{sec:IV-B}.

\subsection{Height and Attitude PID Control Design}
\label{sec:IV-A}
In this framework, both height and attitude regulation are implemented using PID controllers.  
The height controller receives the desired altitude $z^{d}$ and the measured altitude $z$, and computes the desired vertical acceleration $a_z$. Based on the altitude error $e_z = z^{d}-z$ and vertical velocity error $e_{v_z}=-\dot{z}$, the control law is
\begin{equation}
a_z = {K_p}^z e_z + {K_i}^z \int e_z \, dt + {K_d}^z e_{v_z}.
\label{eq:a_z}
\end{equation}
where ${K_p}^z$, ${K_i}^z$, and ${K_d}^z$ denote the proportional, integral, and derivative gains of the altitude PID controller, respectively.

Attitude control follows a similar PID structure with nonlinear attitude-error feedback. Let $\bm{\zeta}^{d}=[\varphi^{d},\theta^{d},\psi^{d}]^{\mathrm{T}}$ and $\bm{\zeta}=[\varphi,\theta,\psi]^{\mathrm{T}}$ be the desired and actual Euler angles. The controller computes the attitude error using the rotation matrices $\bm{R}_{b}^{e}$ and ${\bm{R}_{b}^{e}}^{d}$, forming
\begin{equation}
\bm{\widetilde{R}} = \bm{R}_{b}^{e\mathrm{T}} {\bm{R}_{b}^{e}}^d,
\label{eq:R}
\end{equation}
and extracting the attitude error vector
\begin{equation}
\bm{e}_R \triangleq \frac{1}{2}\,\mathrm{vex}\!\left(\bm{\widetilde{R}}^{\mathrm{T}} - \bm{\widetilde{R}}\right).
\label{eq:e_R}
\end{equation}
where $\mathrm{vex}(\cdot)$ denotes the inverse of the hat operator.
Euler angles are converted to quaternions and then to direction cosine matrices to ensure numerical stability in computing $\bm{e}_R$.

Using the attitude error $\bm{e}_R$ and its rate $\bm{\dot{e}}_R$, the controller generates the desired control moments $\bm{M}^{d}=[{M_x}^{d},{M_y}^{d},{M_z}^{d}]^{\mathrm{T}}$ through
\begin{equation}
\begin{split}
M_i^d &= {K_p}^{\zeta} {e_{R,i}}
      + {K_i}^{\zeta} \int {e_{R,i}} \, dt  \\
      &\quad + {K_d}^{\zeta} {\dot{e}_{R,i}},\quad
      i\in\{x, y, z\},
\end{split}
\label{eq:M_i}
\end{equation}
where ${K_p}^{\zeta}$, ${K_i}^{\zeta}$, and ${K_d}^{\zeta}$ denote the proportional, integral, and derivative gains of the attitude PID controller, respectively. This unified PID design ensures stable altitude tracking and robust attitude regulation under dynamic flight conditions.

\subsection{Design of Hybrid Disturbance Observer}
\label{sec:IV-B}
In this design, the disturbance observer (DOB) estimates disturbances acting on the height and attitude PID controllers, thereby enabling disturbance compensation. For complex systems such as quadrotors, each controlled variable can be approximated as a generic second-order dynamic system, following from \hyperref[eq:dynamics]{(\ref*{eq:dynamics})}:
\begin{equation}
m\bm{a} = \bm{f}_{u} + \bm{f}_{d},
\label{eq:ma}
\end{equation}
\begin{equation}
\bm{I}\dot{\bm{\omega}} + \bm{\omega} \times (\bm{I}\bm{\omega})
= \bm{M}_{u} + \bm{M}_{d},
\label{eq:iw}
\end{equation}
where $\bm{a}$ and $\dot{\bm{\omega}}$ are the linear and angular accelerations estimated via numerical differentiation, the term $\bm{\omega}\times(\bm{I}\bm{\omega})$ denotes the gyroscopic effect. The control inputs are denoted by $\bm{f}_u$ and $\bm{M}_u$, while $\bm{f}_d$ and $\bm{M}_d$ represent disturbances arising from unmodeled dynamics, external forces, and nonlinear coupling.

To indirectly estimate the unknown disturbances defined in \hyperref[eq:ma]{(\ref*{eq:ma})} and \hyperref[eq:iw]{(\ref*{eq:iw})}, the DOB utilizes the discrepancy between the measured acceleration and the acceleration predicted by control inputs. The disturbance estimates are given by:
\begin{equation}
\hat{\bm{f}}_d = m \hat{\bm{a}} - \bm{f}_u,
\label{eq:fd}
\end{equation}
\begin{equation}
\hat{\bm{M}}_d = \bm{I} \hat{\dot{\bm{\omega}}} + \bm{\omega} \times (\bm{I}\bm{\omega}) - \bm{M}_u,
\label{eq:md}
\end{equation}
where $\hat{\bm{f}}_d$ and $\hat{\bm{M}}_d$ denote the estimated disturbance force and moment, and $\hat{\bm{a}}$ and $\hat{\dot{\bm{\omega}}}$ are the corresponding accelerations obtained from sensor data. As direct acceleration measurements are typically unavailable, they are approximated by numerically differentiating linear and angular velocities.

External disturbances are typically non-stationary and time-varying, and numerical differentiation can amplify measurement noise, leading to instability, signal contamination, or degraded control performance. 
Therefore, a robust three-stage filtering scheme is adopted for disturbance estimation, consisting of a Median Filter (MF) for outlier rejection, an Exponentially Weighted Moving Average (EMA) for smoothing, and a first-order Infinite Impulse Response (IIR) low-pass filter for noise attenuation and disturbance tracking, as illustrated in Fig.~\ref{fig:HDOB}. 
This hybrid structure balances noise suppression with disturbance responsiveness. To suppress impulsive noise and disturbances, a median filter is first applied to recent disturbance estimates, with the median value within the sliding window taken as the current output:
\begin{equation}
x_d^{\text{med}}(t)
= \mathrm{Median}\left\{
\hat{x}_d(t-k\Delta t)\,\middle|\,
k=0,1,\ldots,N-1
\right\}.
\label{eq:med}
\end{equation}
where \(x \in \{f_z, M_x, M_y, M_z\}\) denotes a force or moment component, and \(\hat{x}_d(t-k\Delta t)\) denotes the preliminary disturbance estimate at the \(k\)-th sampling instant. 
The window length \(N\) is typically chosen as an odd integer, and the median-filtered disturbance \(x_d^{\text{med}}(t)\) is obtained as the median of the most recent \(N\) estimates, which provides robustness against impulsive noise and extreme outliers.

To prevent anomalous disturbance estimates from propagating into the first-order low-pass filter, a residual gating mechanism is employed. 
The residual is defined as \(r_x(t)=\left|\hat{x}_d(t)-x_d^{\text{med}}(t)\right|\), where \(\hat{x}_d(t)\) is the preliminary disturbance estimate and \(x_d^{\text{med}}(t)\) is its median-filtered counterpart.
If the residual is below a predefined threshold $\delta_x$, the estimate is considered reliable and is used to update the filter; otherwise, the filter state is held constant to preserve the previous output:
\begin{equation}
d_{b}^{x}(t)=
\begin{cases}
d_{b}^{x}(t-1)+\alpha_x\left(\hat{x}_d(t)-d_{b}^{x}(t-1)\right), & r_x(t)\le\delta_x,\\
d_{b}^{x}(t-1), & r_x(t)>\delta_x,
\end{cases}
\label{eq:db}
\end{equation}
where $\alpha_x$ is the filter update coefficient, ${d}_{b}^{x}(t)$ denotes the current output of the first-order low-pass filter, and ${d}_{b}^{x}(t-1)$ denotes the filtered value from the previous time step.

To further improve the smoothness and stability of disturbance estimation, an independent exponential moving average (EMA) filter is applied in parallel with the gated first-order low-pass filter. It smooths the preliminary estimates \(\hat{{x}}_d(t)\) to suppress transient noise and abrupt fluctuations, thereby improving the robustness of disturbance observation:
\begin{equation}
d_{e}^{x}(t)=\beta_x d_{e}^{x}(t-1)+(1-\beta_x)\hat{x}_d(t),
\label{eq:de}
\end{equation}
where \({d}_{e}^{x}(t)\) denotes the EMA-filtered output and \(\beta_x\) is the weighting factor. A larger \(\beta_x\) yields a smoother but slower response, whereas a smaller \(\beta_x\) enables faster adaptation to recent observations. 
By assigning exponentially decaying weights to past disturbance estimates, the EMA filter attenuates high-frequency noise and provides complementary smoothing to the gated low-pass filter.

Building upon the aforementioned two filtering mechanisms from \hyperref[eq:db]{(\ref*{eq:db})} and \hyperref[eq:de]{(\ref*{eq:de})}, the outputs of the gated low-pass filter and the EMA filter are combined through a weighted fusion scheme to obtain the final disturbance estimate:
\begin{equation}
d_{\text{est}}^{x}(t)=\lambda_d d_{b}^{x}(t)+(1-\lambda_d)d_{e}^{x}(t),
\label{eq:dest}
\end{equation}
where \({d}_{\text{est}}^{x}(t)\) denotes the final disturbance estimate and \(\lambda_d\in[0,1]\) is the fusion weight coefficient that balances the contributions of the two filtered outputs. These estimates are incorporated into the height and attitude control loops to enable online compensation of the control inputs.

Using the fused disturbance estimates, the control inputs for the height and attitude channels are compensated. For the \(z\)-axis, the vertical thrust command is adjusted by subtracting the estimated disturbance, enabling the PID-generated acceleration \(a_z\) to achieve the desired altitude while rejecting external disturbances:
\begin{equation}
f_z = m(g + a_z) - d_{\text{est}}^{f_z}.
\label{eq:fz}
\end{equation}

Similarly, in the attitude control loop, external disturbances manifest as disturbance torques through the system's dynamic coupling. The disturbance-compensated angular acceleration about each body axis can be expressed as
\begin{equation}
\dot{\omega}_{i} = \frac{M_{i}^{d}}{I_{i}} - \frac{d_{\text{est}}^{M_i}}{I_{i}},\quad
i\in\{x,y,z\},
\label{eq:w}
\end{equation}
where $d_{\text{est}}^{M_i}$ denotes the estimated disturbance torque acting about the corresponding body axis.

\section{Experiments and Results}
\label{sec:V}
To validate the effectiveness and superiority of the proposed hybrid cascaded control architecture, this section conducts comprehensive comparisons with several advanced methods based on reinforcement learning and classical control. 
The evaluation includes parameter and training configurations, ablation studies, various disturbance tests, and multiple trajectory tracking tasks to assess the performance and stability of the proposed approach.

\subsection{Relevant Parameter Settings}
\label{sec:V-A}
All simulations were conducted within the PyBullet \cite{ref34} physics simulation environment, built on the PyTorch framework. The physical parameters of the Crazyflie 2.0 \cite{ref35} quadrotor used in the simulations are summarized in Table~\ref{tab:crazyflie_params}. 

For horizontal-position control, TD3 was adopted with a PID expert policy, whose gains were set to $K_{p}^{E}=2.0$, $K_{i}^{E}=0.0$, and $K_{d}^{E}=0.5$. The expert policy network used two hidden layers of 128 neurons and was trained with a learning rate of $5 \times 10^{-4}$. 
Both actor and critic networks used hidden layers of 64 neurons, and the output actions were constrained to $[-1,1]$ by a Tanh activation because of thrust limitations. The actor and critic learning rates were initialized as $l_a=5\times10^{-4}$ and $l_c=10^{-3}$, kept unchanged for the first 100 episodes, 
and then linearly decayed over about 67 episodes to $l_a=10^{-4}$ and $l_c=10^{-5}$, after which they were fixed until the end of the 200-episode training process. The remaining hyperparameters and control parameters are listed in Table~\ref{tab:params}. All comparative experiments used identical controller and observer settings with the same random seeds.

\begin{table}[!t]
    \scriptsize
    \setlength{\tabcolsep}{6pt}
    \centering
    \caption{Relevant Physical Parameters Used in Simulation}
    \label{tab:crazyflie_params}
    \begin{tabular}{llll}
        \toprule
        Description & Symbol & Value & Unit \\
        \midrule
        Mass & $m$ & 0.027 & \si{kg} \\
        Gravitational acceleration & $g$ & 9.81 & \si{m/s^2} \\
        Quadrotor arm length & $l$ & 0.0397 & \si{m} \\
        Moment of inertia (x-axis) & $I_{x}$ & $1.40 \times 10^{-5}$ & \si{kg.m^2} \\
        Moment of inertia (y-axis) & $I_{y}$ & $1.40 \times 10^{-5}$ & \si{kg.m^2} \\
        Moment of inertia (z-axis) & $I_{z}$ & $2.17 \times 10^{-5}$ & \si{kg.m^2} \\
        Lift coefficient & $C_{f}$ & $2.88 \times 10^{-8}$ & \si{kg.m/rad^2} \\
        Torque coefficient & $C_{M}$ & $7.24 \times 10^{-10}$ & \si{kg.m^2/rad^2} \\
        \bottomrule
    \end{tabular}
\end{table}

\begin{table}[!t]
\caption{Configuration Parameters for Simulation and Learning}
\label{tab:params}
\centering
\scriptsize
\setlength{\tabcolsep}{3pt} 
\begin{tabular}{ll}
\toprule
Symbol & Value \\
\midrule
Soft update rate, $\tau$ & $5 \times 10^{-3}$ \\
Discount factor, $\gamma$ & 0.99 \\
Policy update delay, $d$ & 2 \\
Capacities $C_{\mathrm{m}}$, $C_{\mathrm{h}}$ and batch size $B$ & 50000, 25000, 256 \\
Values of $L$, $L_m$, $\rho$, $\epsilon$ & 5000, 1000, 0.9, 0.001 \\
Initial and maximum values of $\xi$ & 0, 1.0 \\
Values of $t_0$, $\beta$, $\kappa$, $k$ & 75000, 0.5, 20000, 5.0 \\
Values of $\alpha_{\min}$, $\alpha_{\max}$, $\sigma_{\mathrm{th}}$ & 0.3, 0.7, 1.5 \\
Reward parameters $\beta_1$, $\beta_2$, $\beta_3$ & 2.0, 0.2, 5.0 \\
Height PID gains, ${K_p}^z$, ${K_i}^z$, ${K_d}^z$ & $3.0$, $0.0$, $2.0$ \\
Roll PID gains, ${K_p}^{\varphi}$, ${K_i}^{\varphi}$, ${K_d}^{\varphi}$ & $10.4$, $0.0$, $1.2$ \\
Pitch PID gains, ${K_p}^{\theta}$, ${K_i}^{\theta}$, ${K_d}^{\theta}$ & $10.4$, $0.0$, $1.2$ \\
Yaw PID gains, ${K_p}^{\psi}$, ${K_i}^{\psi}$, ${K_d}^{\psi}$ & $10.4$, $0.0$, $1.2$ \\
Predefined thresholds, $\delta_{f_z}$, $\delta_{M_x}$, $\delta_{M_y}$, $\delta_{M_z}$ & $0.4$, $0.01$, $0.01$, $0.01$ \\
Update coefficients, $\alpha_{f_z}$, $\alpha_{M_x}$, $\alpha_{M_y}$, $\alpha_{M_z}$ & $0.24$, $0.4$, $0.4$, $0.4$ \\
Weighting factors, $\beta_{f_z}$, $\beta_{M_x}$, $\beta_{M_y}$, $\beta_{M_z}$ & $0.96$, $0.955$, $0.955$, $0.955$ \\
Fusion weight coefficient, $\lambda_d$ & $0.45$ \\
\bottomrule
\end{tabular}
\end{table}

\begin{figure*}[!t]
\centering
\vspace{-0.5cm}
\subfloat[]{\includegraphics[width=0.32\textwidth]{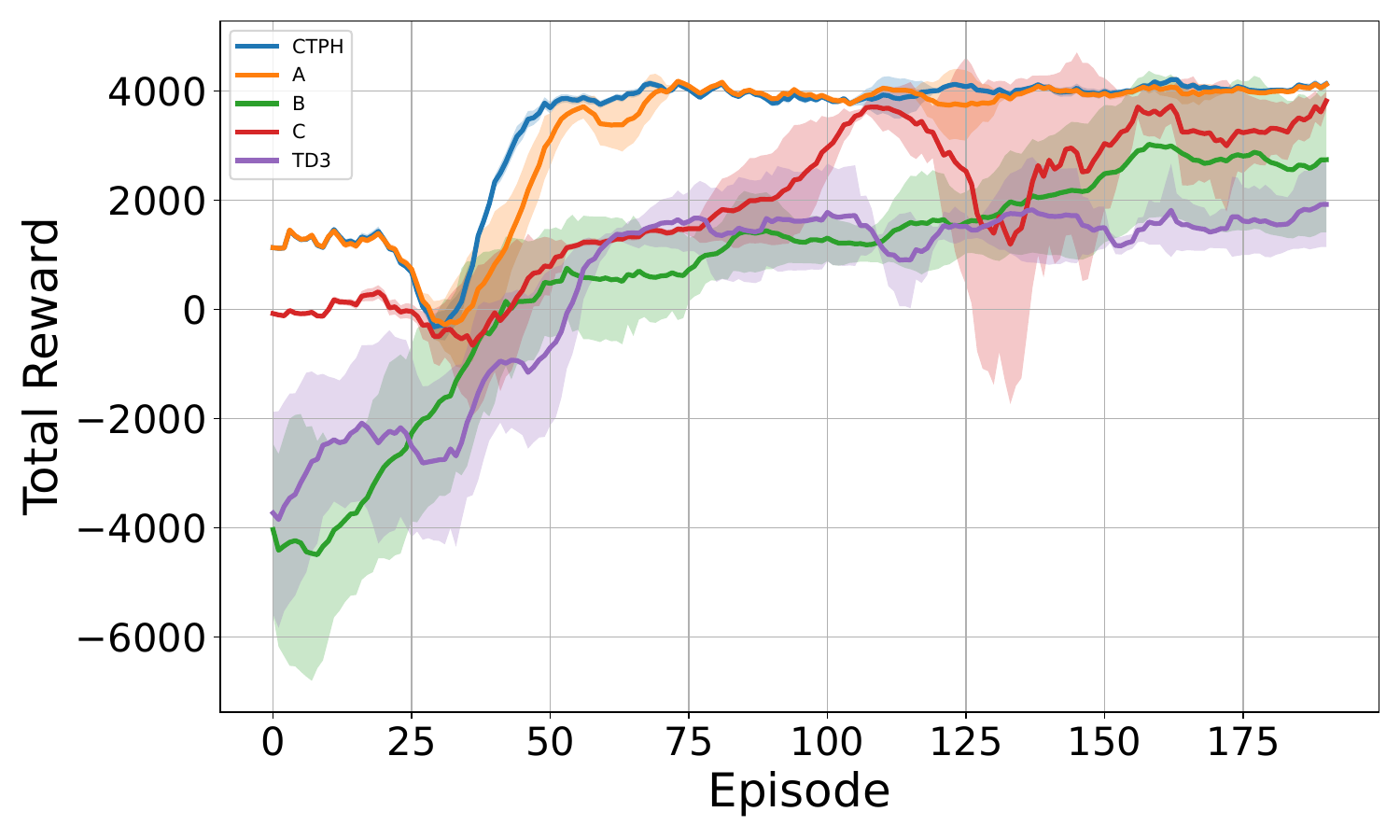}\label{fig:sub1}}
\hfill
\subfloat[]{\includegraphics[width=0.32\textwidth]{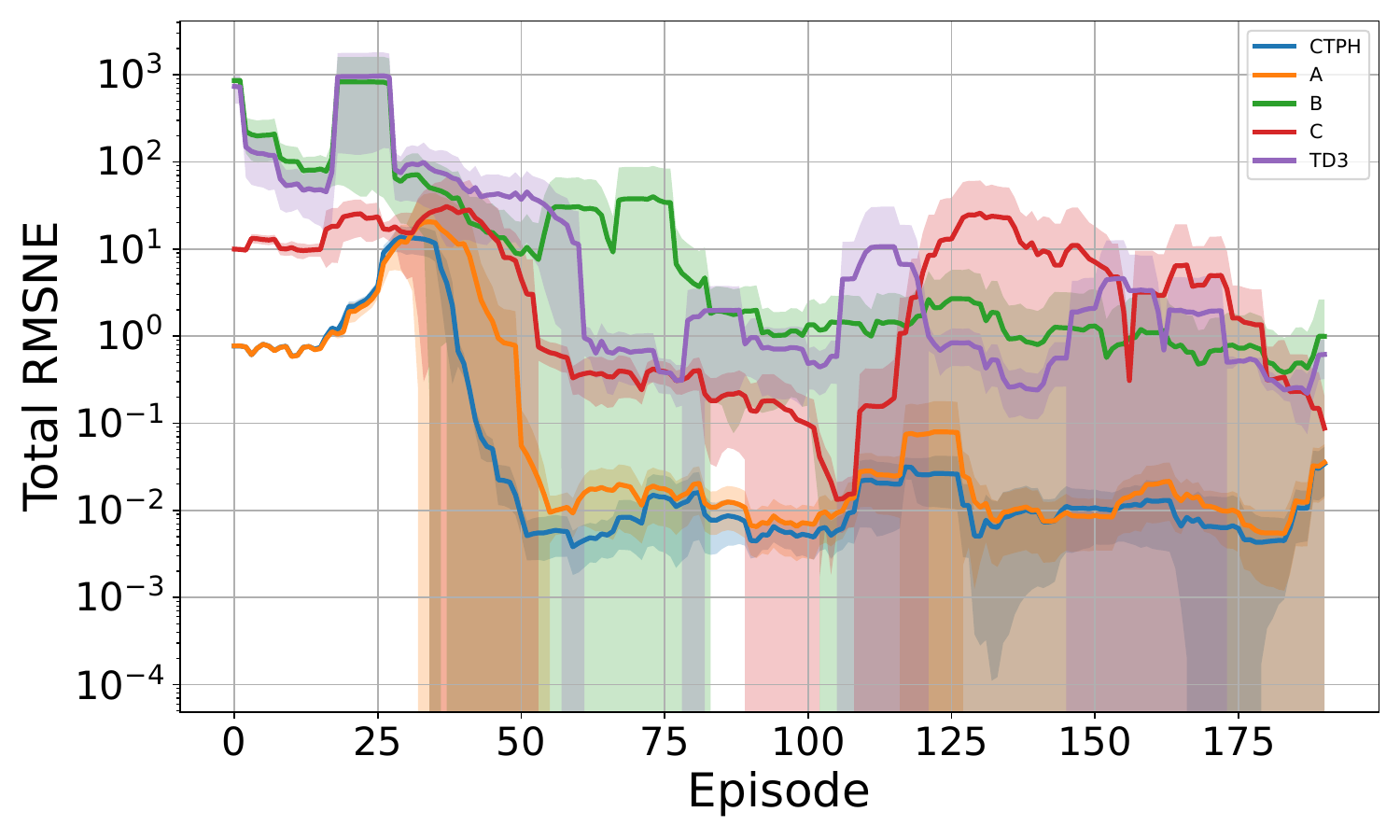}\label{fig:sub2}}
\hfill
\subfloat[]{\includegraphics[width=0.32\textwidth]{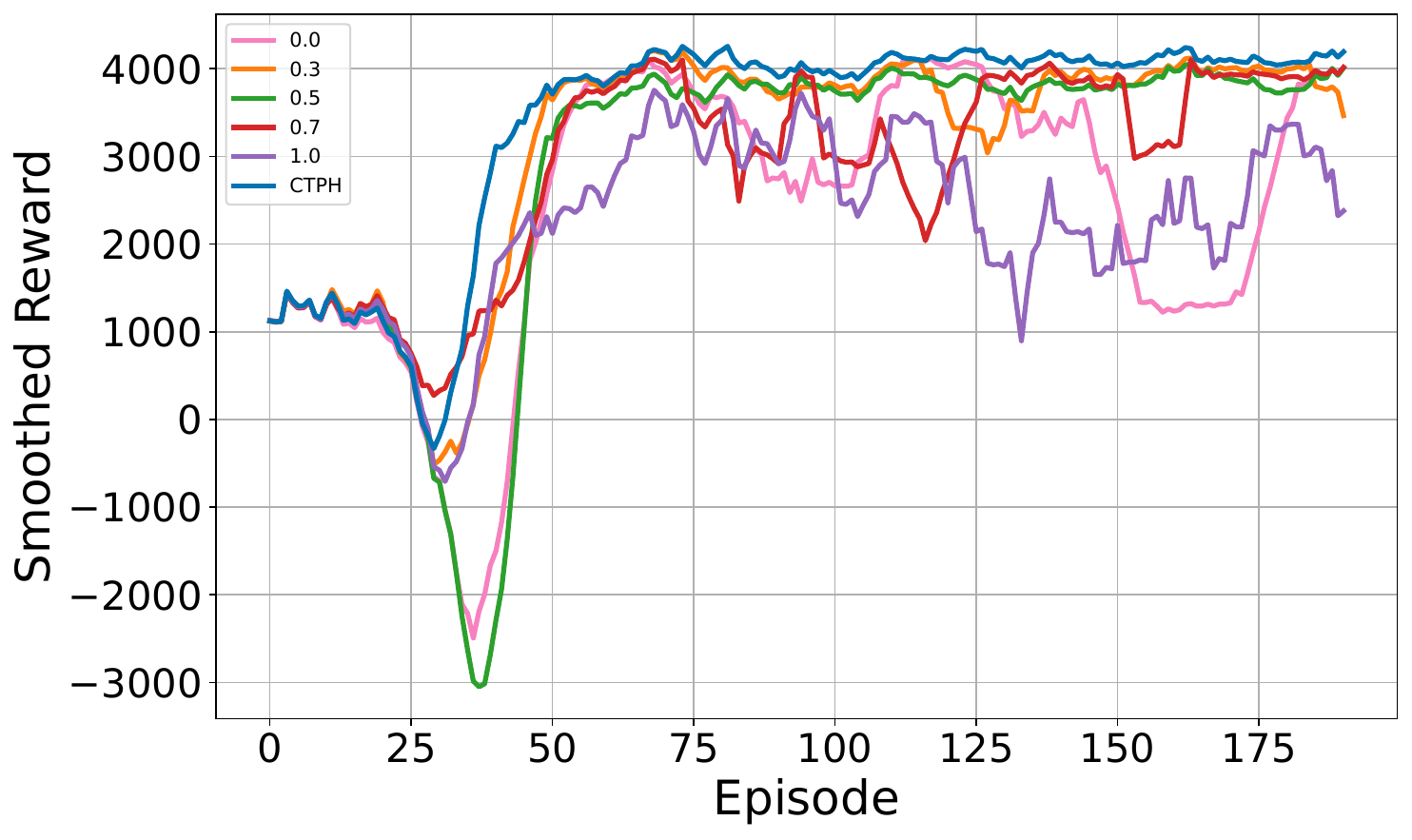}\label{fig:sub3}}
\caption{The training results of the horizontal position TD3 controller are evaluated using the reward and the Root Mean Square Normalized Error (RMSNE) as performance metrics of the agent. (a) Average rewards per episode. (b) Average RMSNE per episode. (c) Average reward curves under different $\alpha$.}
\label{fig:all}
\end{figure*}

\subsection{Training Settings and Results}
\label{sec:V-B}
During training of the horizontal-position TD3 controller, the agent's initial position was randomly sampled from \([-5,5]\), while all initial attitude angles were set to zero. 
The simulation time step was fixed at \(0.01\) s, and both the target position and desired attitude were set to zero. Since the reward mainly depended on the horizontal displacement from the target, 
the objective was to drive the agent to the target coordinate as efficiently as possible within each episode. Training was conducted for 200 episodes, each limited to 750 time steps; once the limit was reached, the episode was terminated and restarted from a newly sampled initial position. 
To analyze the control strategy, an ablation study was conducted to examine the effects of different improvement modules of the horizontal-position TD3 controller under identical environmental settings. The following five variants were trained and evaluated:
\begin{enumerate}
    \item Proposed Method (CTPH): the full method with all three proposed enhancements.
    \item Without Dual Experience Replay (A): CTPH without the dual experience replay buffer.
    \item Without PID-Based Expert Policy (B): CTPH without the PID-based expert guidance module.
    \item Without Aggregated Q-Network Architecture (C): CTPH without the aggregated Q-network architecture.
    \item Baseline TD3 (TD3): the standard TD3 algorithm used as the ablation baseline.
\end{enumerate}

As illustrated in Fig.~\ref{fig:all}\subref{fig:sub1}, the cumulative reward trends of different TD3-based variants are compared under the same environment setting. The solid lines denote the mean smoothed returns over five independent trials, and the shaded regions indicate the corresponding standard deviations. 
At the early stage of training, higher rewards are achieved by CTPH, A, and C because the expert guidance module is included. As training proceeds, however, method C becomes more prone to unstable convergence and inconsistent learning in the absence of the aggregated Q-network architecture. 
Compared with method A, CTPH achieves stable rewards earlier through reward-filtered dual experience replay and also shows smaller performance variations across trials. In addition, incorporating individual components in isolation does not necessarily yield satisfactory performance: 
compared with TD3, method B performs poorly in early training, likely because low-quality samples are introduced into the high-reward buffer without expert guidance.

Due to the stochasticity of the initial positions during training, the Root Mean Squared Normalized Error (RMSNE) is used to evaluate both the final performance of the learned horizontal-position TD3 controller and its convergence behavior. 
This metric normalizes the error by the initial tracking error and is computed over the latter half of each episode, thereby better reflecting steady-state performance. 
Let $e_t$ denote the scalar tracking error at discrete time step $t$, and let $e_i$ denote the initial error at the beginning of the episode. Given an episode of $N$ time steps, RMSNE is computed over the latter half of the episode, i.e., for $t = N/2, \ldots, N$, as:
\begin{equation}
\text{RMSNE} = \sqrt{\frac{2}{N} \sum_{t=N/2}^{N} \left( \frac{e_t}{e_i + \epsilon_0} \right)^2},
\label{eq:RMSNE}
\end{equation}
where $\epsilon_0$ is a small positive constant introduced to prevent division by zero and to ensure numerical stability.
As shown in Fig.~\ref{fig:all}\subref{fig:sub2}, the RMSNE-based position tracking performance indicates that the proposed method achieves faster learning and better convergence compared to the other approaches.

To further evaluate the proposed aggregated Q-network with adaptive weighting, simulations were conducted by comparing agents trained with five fixed weighting factors (\(\alpha=0.0, 0.3, 0.5, 0.7, 1.0\)) against the proposed adaptive-\(\alpha\) method. 
As shown in Fig.~\ref{fig:all}\subref{fig:sub3}, the adaptive-\(\alpha\) agent achieved higher rewards than the fixed-weight agents. Among fixed settings, moderate values (\(\alpha=0.3\) and \(\alpha=0.5\)) outperformed extreme values, highlighting the importance of balancing the minimum and average of the three Q-values. 
Nonetheless, fixed strategies remained less effective than the adaptive approach, where \(\alpha\) was dynamically adjusted during training. This dynamic adjustment better mitigated underestimation and overestimation biases at learning stages, improving policy stability and efficiency. 
These results confirm the advantage of the proposed aggregated Q-network with adaptive weighting.

\begin{figure}[!t]
\centering
\subfloat[]{\includegraphics[width=0.90\columnwidth]{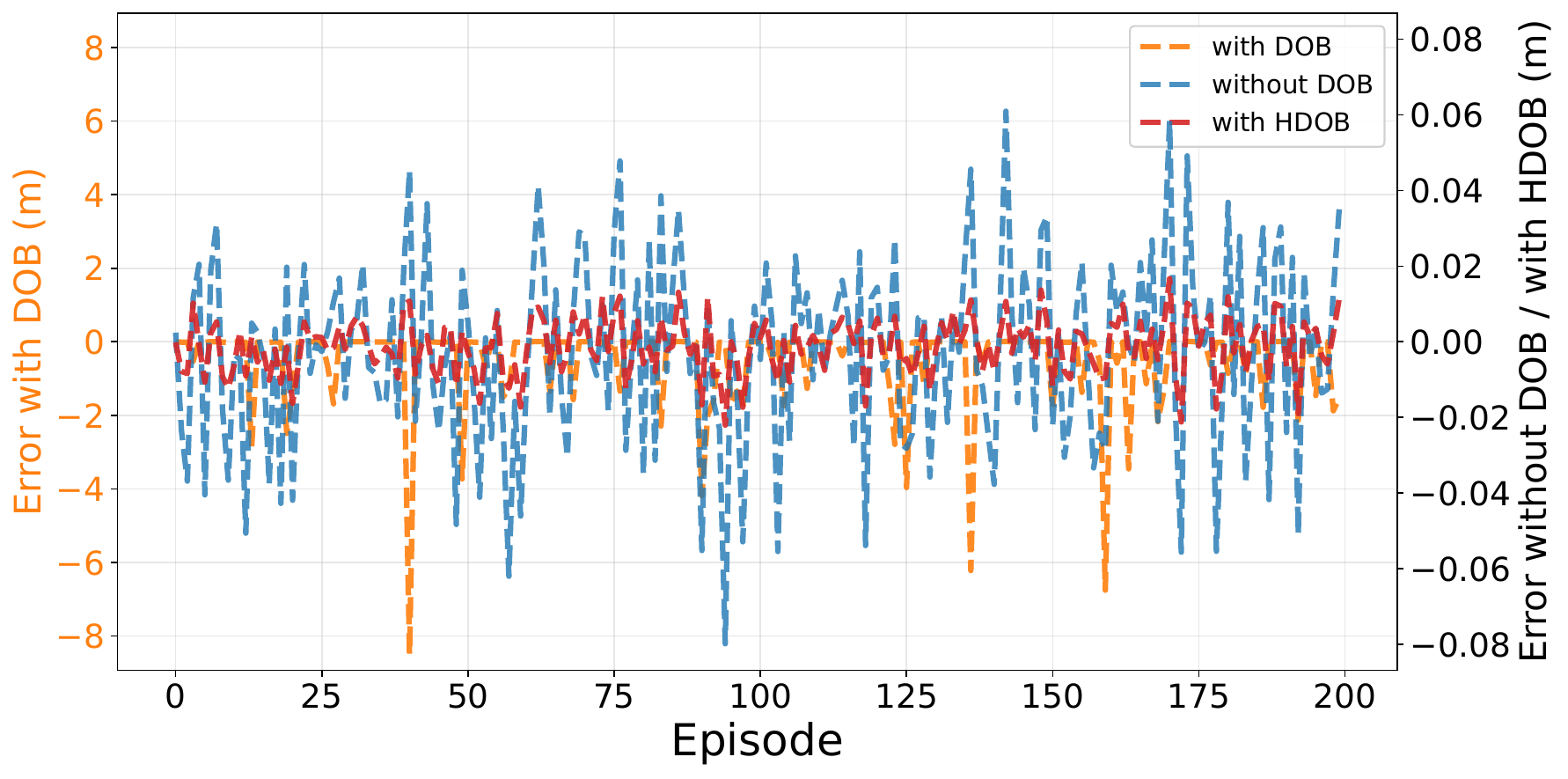}\label{fig:sub4}}
\\[0.3em] 
\subfloat[]{\includegraphics[width=0.90\columnwidth]{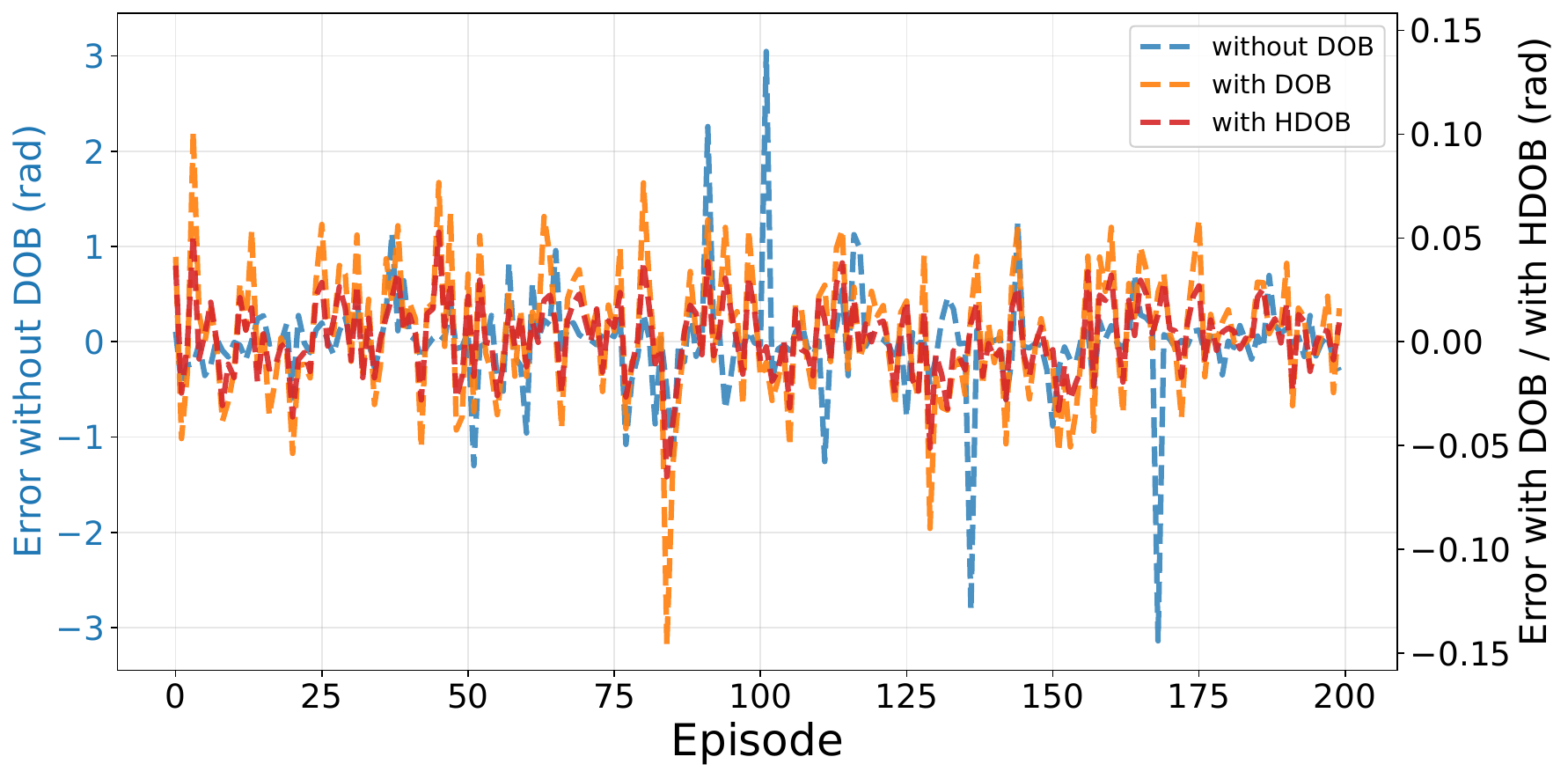}\label{fig:sub5}}
\caption{Comparison of final errors over 200 episodes for the PID controller without DOB, with baseline DOB, and with the proposed HDOB. (a) Performance of the height PID controller. (b) Performance of the yaw PID controller.}
\label{fig:overall}
\end{figure}

\begin{figure}[!t]
    \centering
    \includegraphics[width=0.6\linewidth]{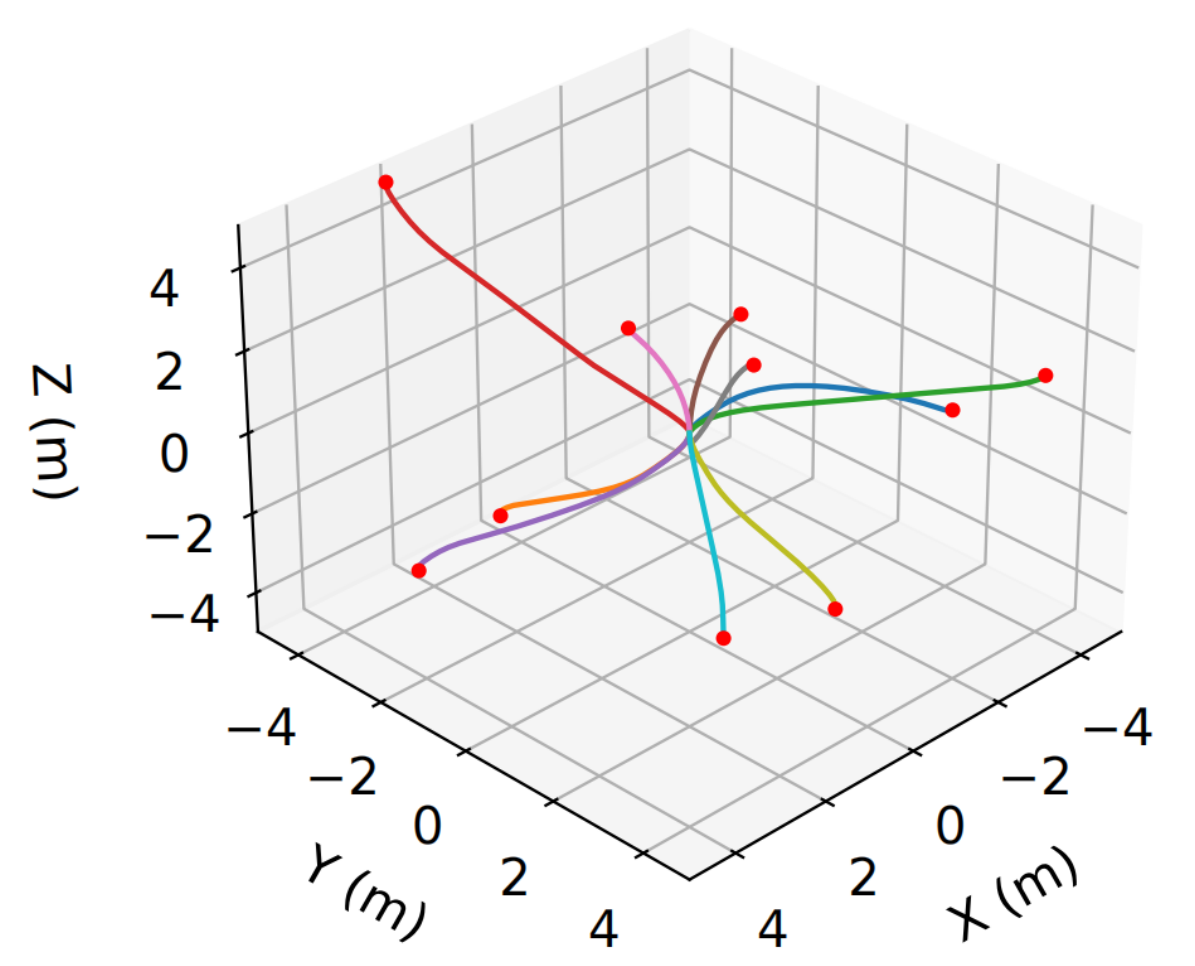}
    \caption{3D flight trajectories of the quadrotor toward 10 random target points}
    \label{fig:3D}
\end{figure}

\begin{table}[!t]
    \scriptsize  
    \renewcommand{\arraystretch}{0.95}  
    \setlength{\tabcolsep}{3pt} 
    \centering
    \caption{Control Performance Statistics for Height and Yaw. w/o DOB: without DOB; w/ DOB: with DOB; w/ HDOB: with HDOB.}
    \label{tab:Control_error_combined}
    \begin{tabular}{llcccc}
        \toprule
        Control & Method & MAE & MAX & MIN & STD \\
        \midrule
        \multirow{3}{*}{\shortstack{Height \\ (m)}} 
            & w/o DOB & 0.0199 & 0.0798 & 0.0002 & 0.0153 \\
            & w/ DOB  & 0.4774 & 8.5023 & $3.80\times10^{-5}$ & 1.155 \\
            & w/ HDOB & \textbf{0.0061} & \textbf{0.0220} & $\boldsymbol{2.27\times10^{-5}}$ & \textbf{0.0044} \\
        \midrule
        \multirow{3}{*}{\shortstack{Yaw \\ (rad)}} 
            & w/o DOB & 0.2732 & 3.1390 & 0.0056 & 0.4589 \\
            & w/ DOB  & 0.0243 & 0.1459 & 0.0001 & 0.0204 \\
            & w/ HDOB & \textbf{0.0150} & \textbf{0.0650} & $\boldsymbol{1.52\times10^{-5}}$ & \textbf{0.0113} \\
        \bottomrule
    \end{tabular}
\end{table}

\subsection{Evaluation of HDOB Performance}
\label{sec:V-C}
To simulate dynamic environmental variations, a randomized initial flight state selection strategy and external disturbances are incorporated into the numerical system to assess the effectiveness of the proposed HDOB.
Independent tests were conducted on the attitude and altitude PID controllers. Considering the symmetry of quadrotor attitude dynamics, the yaw PID controller and the altitude PID controller were selected as representative cases for evaluation. In each test, the initial yaw angle was randomly initialized within the range $[-\pi, \pi]$, 
and the initial altitude was randomly selected from $[-5, 5]$, with both the desired yaw angle and altitude set to zero. For comparison, two baseline controllers were considered: a conventional PID controller without any disturbance rejection mechanism, and a PID controller integrated with a baseline DOB only.

To rigorously assess the robustness of the attitude and altitude PID controllers, two distinct categories of complex disturbances are independently introduced into the testing environment:
\begin{equation}
\begin{aligned}
d_z(t) &= I_z(t) + s\left(0.05 + 0.05 \sin(\pi t) + \mathcal{N}(0, 0.05^2)\right), \\
d_{\psi}(t) &= I_{\psi}(t) + s\left(0.0005 + 0.0005 \sin(\pi t)\right) \\
           &\quad + s\,\mathcal{N}(0, 0.0002^2),
\end{aligned}
\label{eq:disturbance}
\end{equation}
where $I_z(t)$ and $I_{\psi}(t)$ denote random pulse disturbances applied with probability 0.5, and $s = \pm 1$ represents a random sign factor.
The external disturbances are considered moderate perturbations and are designed to test the robustness of the PID controllers within typical operational ranges.

\begin{figure*}[!t]
\centering
\vspace{-0.5cm}
\subfloat{\includegraphics[width=0.245\textwidth]{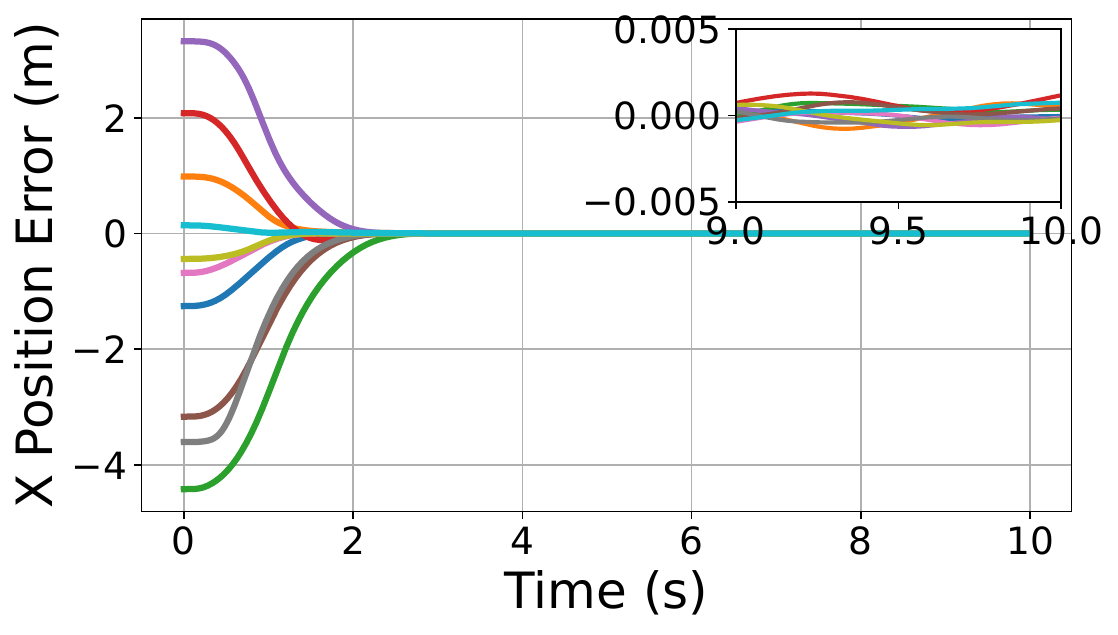}\label{fig:x_pos}}
\subfloat{\includegraphics[width=0.245\textwidth]{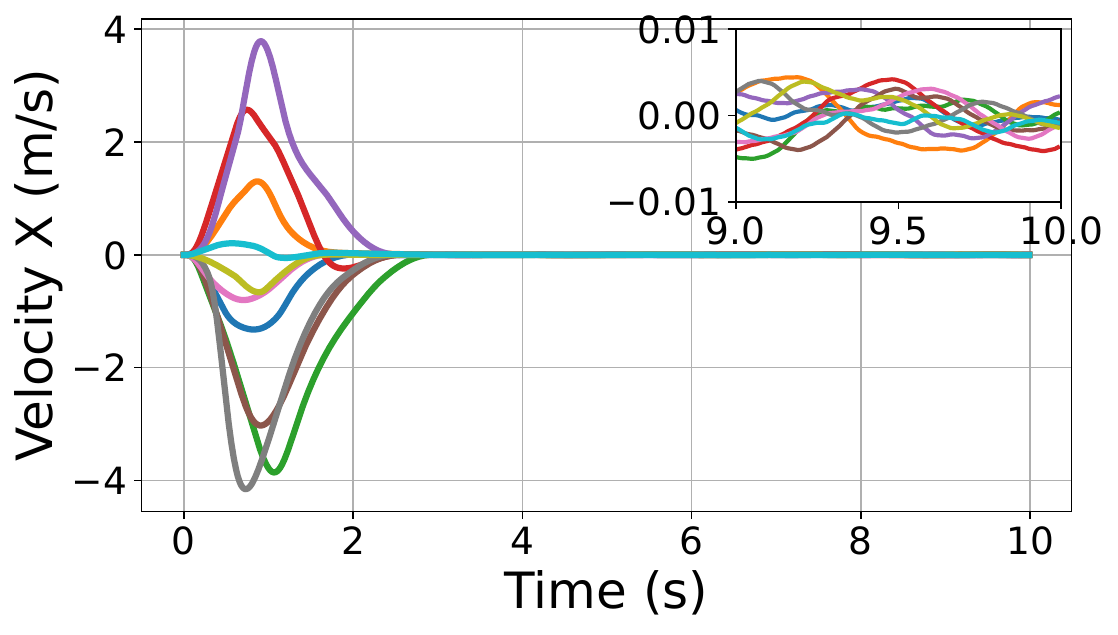}\label{fig:x_vel}}
\subfloat{\includegraphics[width=0.245\textwidth]{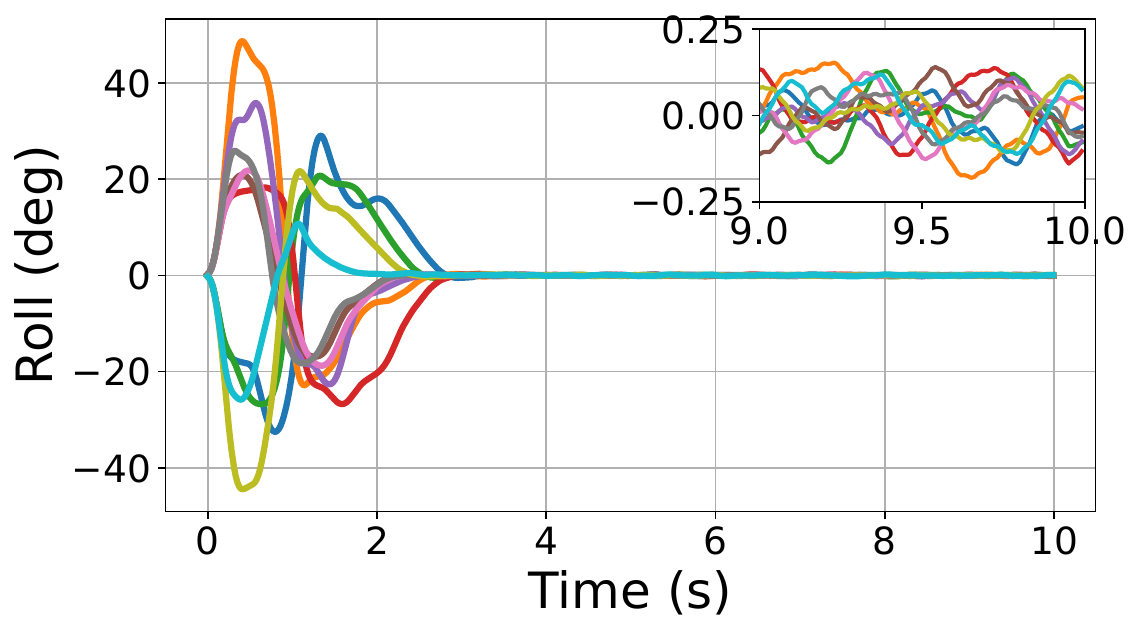}\label{fig:roll}}
\subfloat{\includegraphics[width=0.245\textwidth]{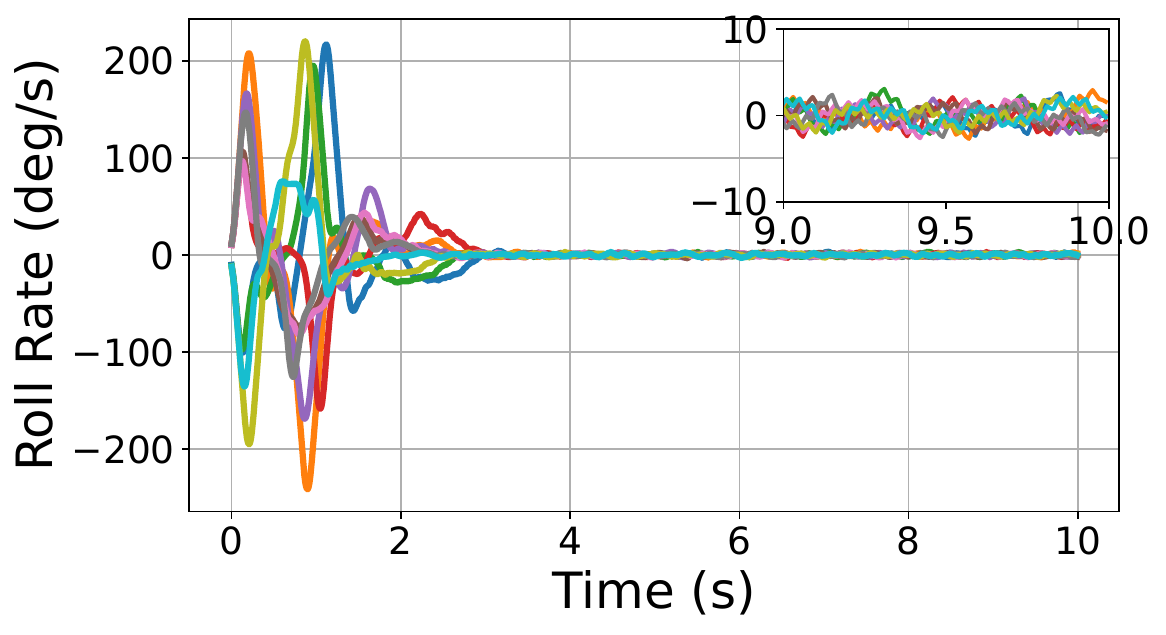}\label{fig:roll_rate}}

\subfloat{\includegraphics[width=0.245\textwidth]{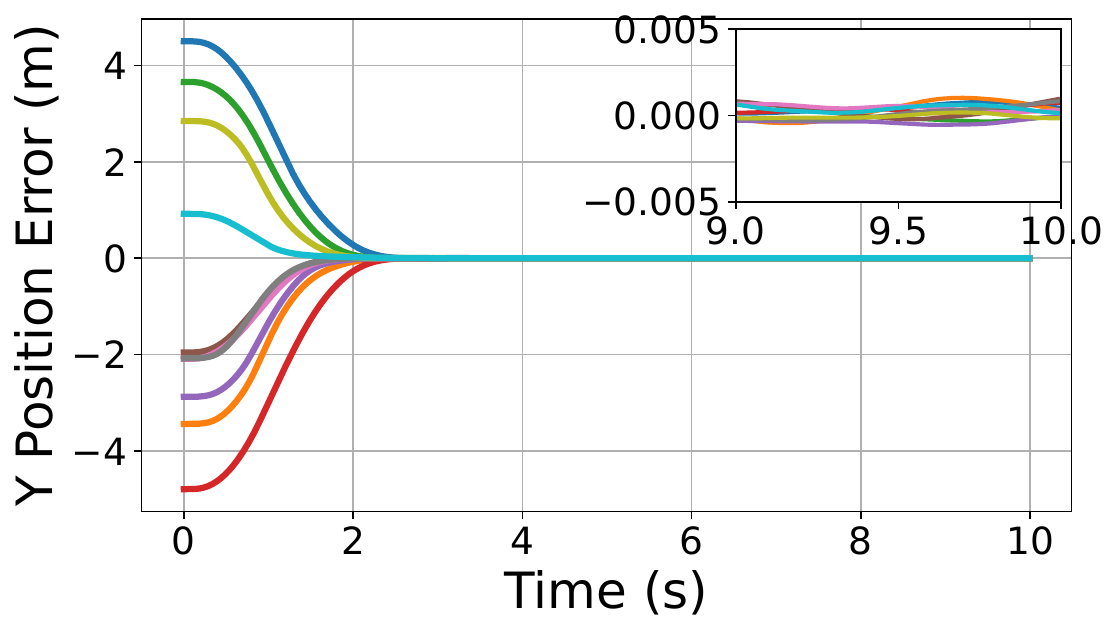}\label{fig:y_pos}}
\subfloat{\includegraphics[width=0.245\textwidth]{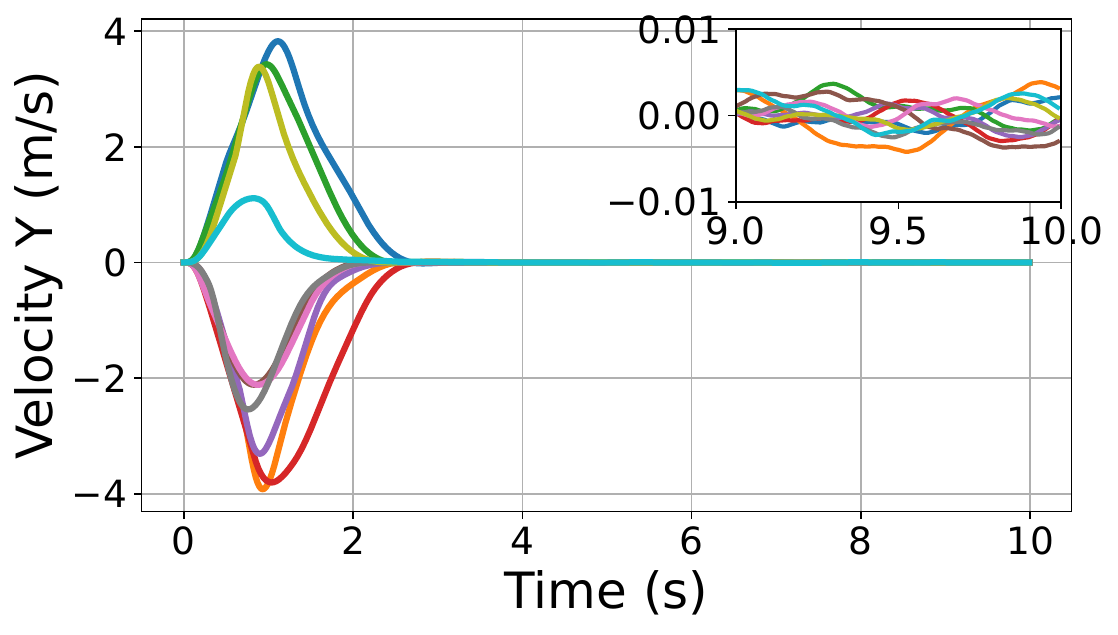}\label{fig:y_vel}}
\subfloat{\includegraphics[width=0.245\textwidth]{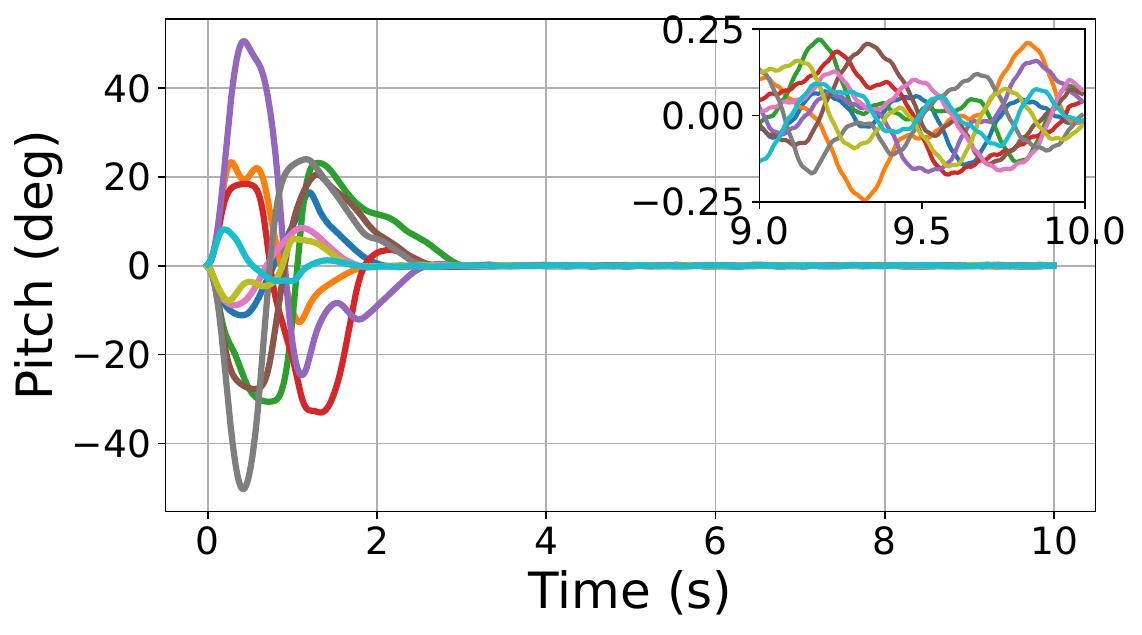}\label{fig:pitch}}
\subfloat{\includegraphics[width=0.245\textwidth]{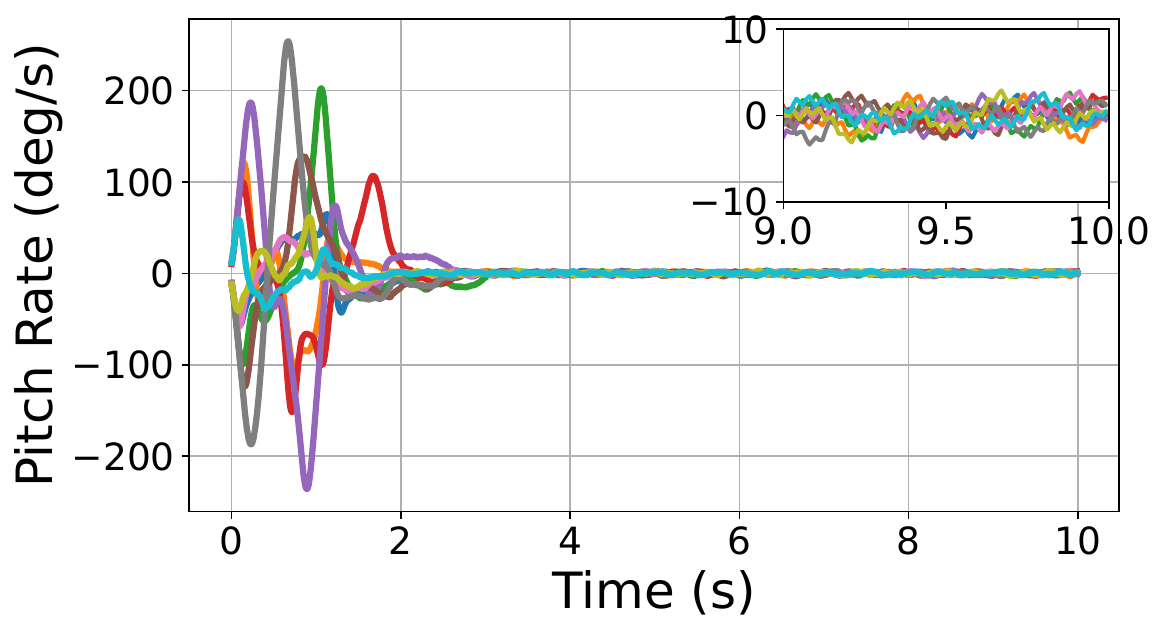}\label{fig:pitch_rate}}

\subfloat{\includegraphics[width=0.245\textwidth]{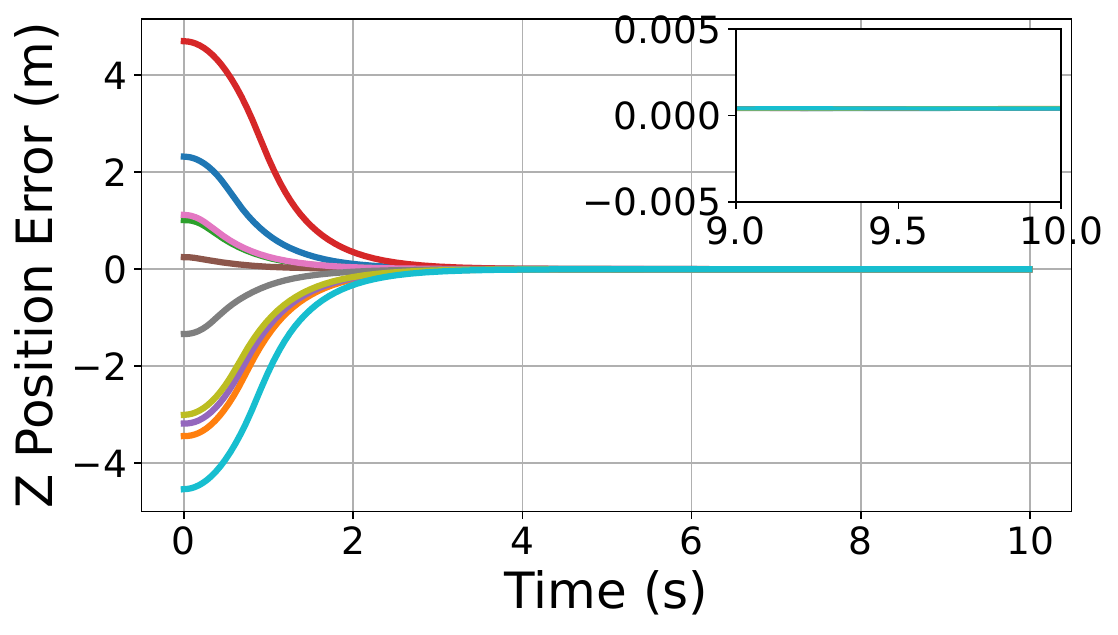}\label{fig:z_pos}}
\subfloat{\includegraphics[width=0.245\textwidth]{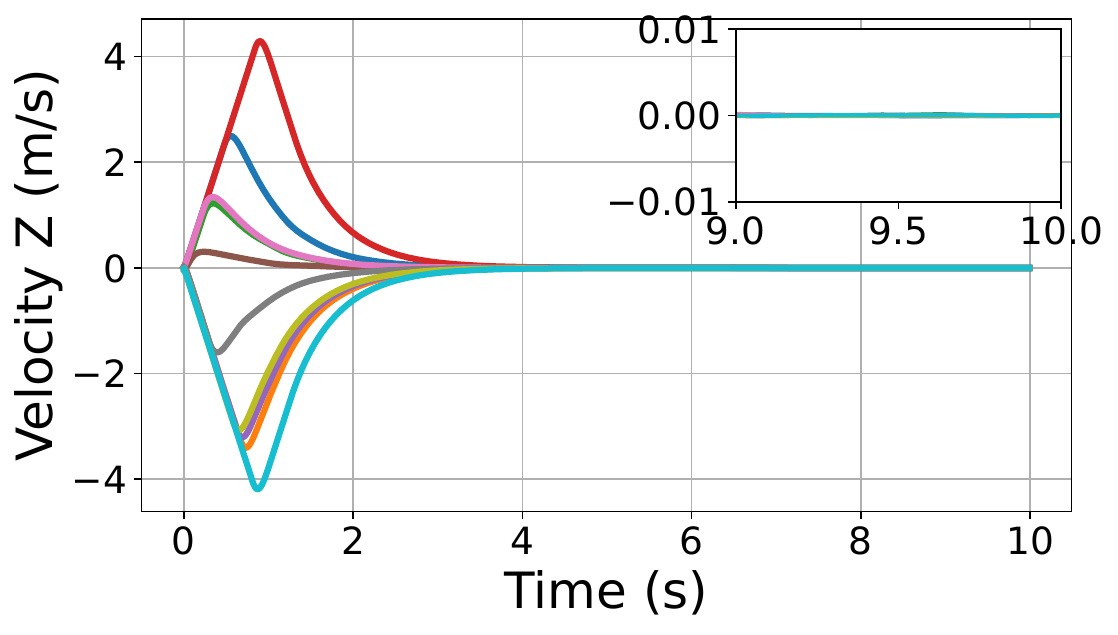}\label{fig:z_vel}}
\subfloat{\includegraphics[width=0.245\textwidth]{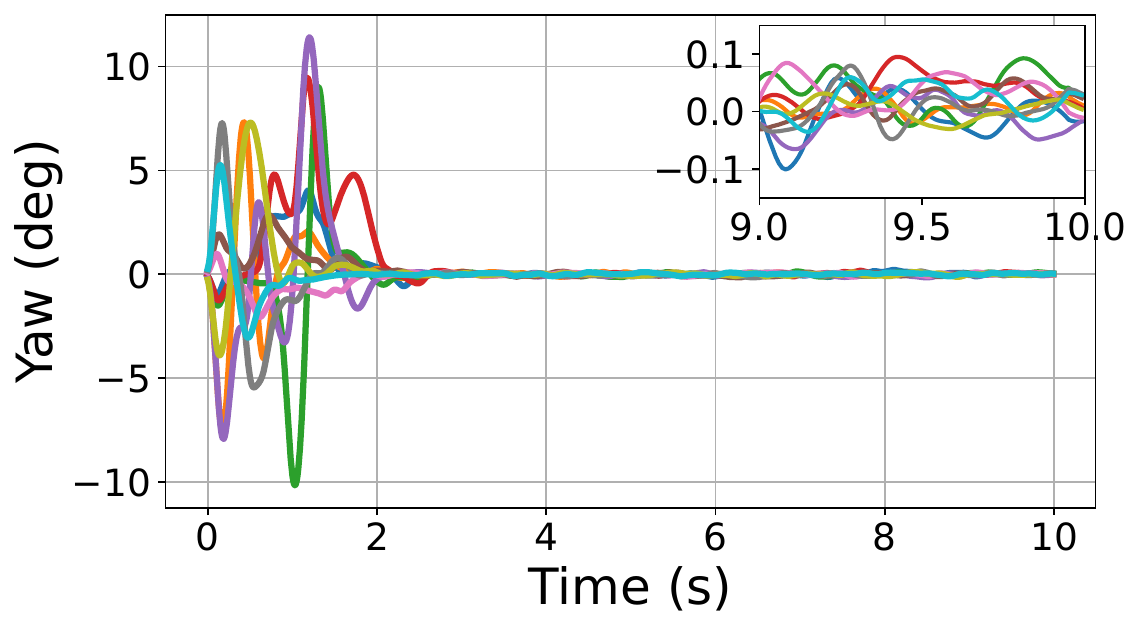}\label{fig:yaw}}
\subfloat{\includegraphics[width=0.245\textwidth]{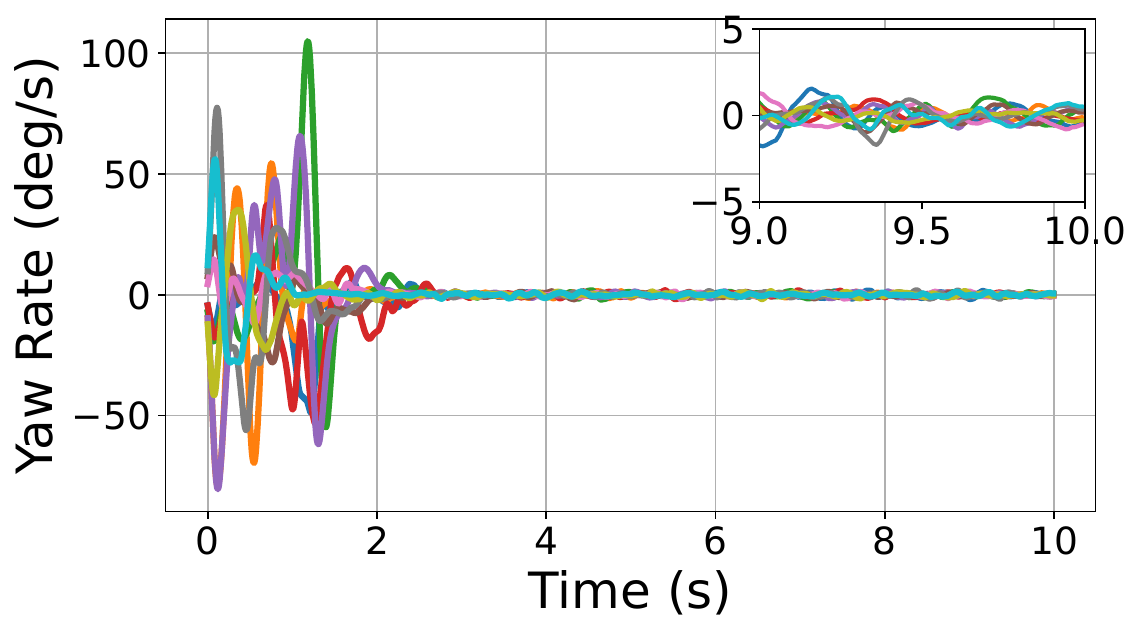}\label{fig:yaw_rate}}
\caption{From left to right and top to bottom: position errors, velocities, attitudes, and angular velocities during the point-to-point trajectory test.}
\label{fig:all_pictures}
\end{figure*}

The final errors over $200$ episodes for the PID controller under altitude control \subref{fig:sub4} and yaw angle control \subref{fig:sub5} tasks are shown in Fig.~\ref{fig:overall}.
Noticeable oscillations were observed in the PID controller's response when subjected to compound disturbances. With the introduction of the baseline DOB in the yaw control loop, partial disturbance mitigation was achieved.
However, fluctuations induced by abrupt perturbation components remained evident. In contrast, the proposed HDOB further suppressed the oscillation peaks and demonstrated a markedly improved disturbance-rejection capability compared with the baseline DOB.
For altitude control, the baseline DOB exhibited pronounced estimation errors due to the presence of high-magnitude transient variations in the disturbance signal, whereas the HDOB maintained accurate disturbance tracking and compensation throughout the test. 

To provide a quantitative comparison, Table~\ref{tab:Control_error_combined} summarizes the final mean absolute error (MAE), maximum error (MAX), minimum error (MIN), and standard deviation (STD) for both the height and yaw PID controllers under different testing conditions.
Together with the preceding analyses, these results indicate enhanced robustness and improved disturbance-rejection performance enabled by the proposed HDOB.

\begin{table}[!t]
    \scriptsize
    \setlength{\tabcolsep}{6pt}
    \centering
    \caption{Point-to-point Flight Tests Performance. All Three Performance Metrics Averaged over 10 Random Trajectories}
    \label{tab:points}
    \begin{tabular}{lccc} 
        \toprule
         & Average SSE (m) & Average Rise Time (s) & Average OS (\%) \\ 
        \midrule
        x & ${3.39\times10^{-4}}$ & 0.93 & 0.78 \\
        y & ${4.08\times10^{-4}}$ & 1.04 & 0.05 \\
        z & ${4.08\times10^{-4}}$ & 1.31 & 0.01 \\
        \bottomrule
    \end{tabular}
\end{table}

\subsection{Random Point-to-Point Flight Tests}
\label{sec:V-D}
To evaluate the effectiveness and stability of the proposed CTPH strategy, 10 target positions \(P^{d}=[x_{R}^{d}, y_{R}^{d}, z_{R}^{d}]\) were randomly generated in a three-dimensional space, with each coordinate uniformly sampled from \([-5,5]\). 
The initial position and attitude angles of the quadrotor were set to zero. The objective was to drive the quadrotor from the initial state to each target stably and rapidly. Each trajectory lasted 1000 time steps with a step size of 0.01\,s. 
As shown in Fig.~\ref{fig:3D}, point-to-point trajectories in 3D space were accurately completed for all predefined random targets. Fig.~\ref{fig:all_pictures} further shows the position errors, velocities, attitudes, and angular velocities along the three axes. 
The proposed CTPH controller reaches the desired states within about 250 time steps, with the position errors approaching zero, while stable attitude and velocity responses are maintained even under rapid velocity changes, indicating strong robustness.

To more intuitively evaluate the performance of point-to-point trajectory tracking, we introduce the above three fundamental performance metrics as additional quantitative validation.
They are defined as:
\begin{equation}
\begin{aligned}
\bar{e}_{\eta} &= \frac{1}{N} \sum_{i=1}^{N} \left( \frac{1}{|S|} \sum_{k \in S} \vert P_{i, \eta}(k) - P_{i, \eta}^{d} \vert \right),\\
\bar{r}_{\eta} &= \frac{1}{N} \sum_{i=1}^{N} \left( t_{90, \eta}^{(i)} - t_{10, \eta}^{(i)} \right),\\
\bar{p}_{\eta} &= \frac{1}{N} \sum_{i=1}^{N} \left( \frac{\max_{k}\left| P_{i, \eta}(k) \right| - \left| P_{i, \eta}^{d} \right|}{\left| P_{i, \eta}^{d} \right| + \varepsilon_p} \right).
\end{aligned}
\label{eq:eta_metrics}
\end{equation}
where $\bar{e}_{\eta}$, $\bar{r}_{\eta}$, and $\bar{p}_{\eta}$ denote the average steady-state error (SSE), average rise time, and average overshoot (OS) along the $\eta$ (\(\eta \in \{x,y,z\}\)) direction, respectively, averaged over \(N=10\) trajectories to reflect the overall tracking performance, and $\varepsilon_p$ is a small positive constant.
Here, $S$ is the index set used to compute the steady-state error over the final time steps, $P_{i,\eta}$ and $P_{i,\eta}^{d}$ denote the current and target positions along the $\eta$-axis in the \(i\)-th experiment, and $t_{90,\eta}^{(i)}$ and $t_{10,\eta}^{(i)}$ represent the time instants at which the $\eta$-axis response in the \(i\)-th trial reaches 90\% and 10\% of the target value.

The numerical results of the point-to-point tracking experiments are presented in Table~\ref{tab:points}. The mean steady-state errors remain close to zero, 
the rise times along all three axes are below \(1.4\) s, and the overshoot remains below \(1\%\), with that along the \(y\)- and \(z\)-axes being practically negligible. These results demonstrate that the proposed CTPH controller achieves fast, accurate, and stable point-to-point tracking.

\begin{figure*}[!t]
\centering
\subfloat{\includegraphics[width=0.22\textwidth]{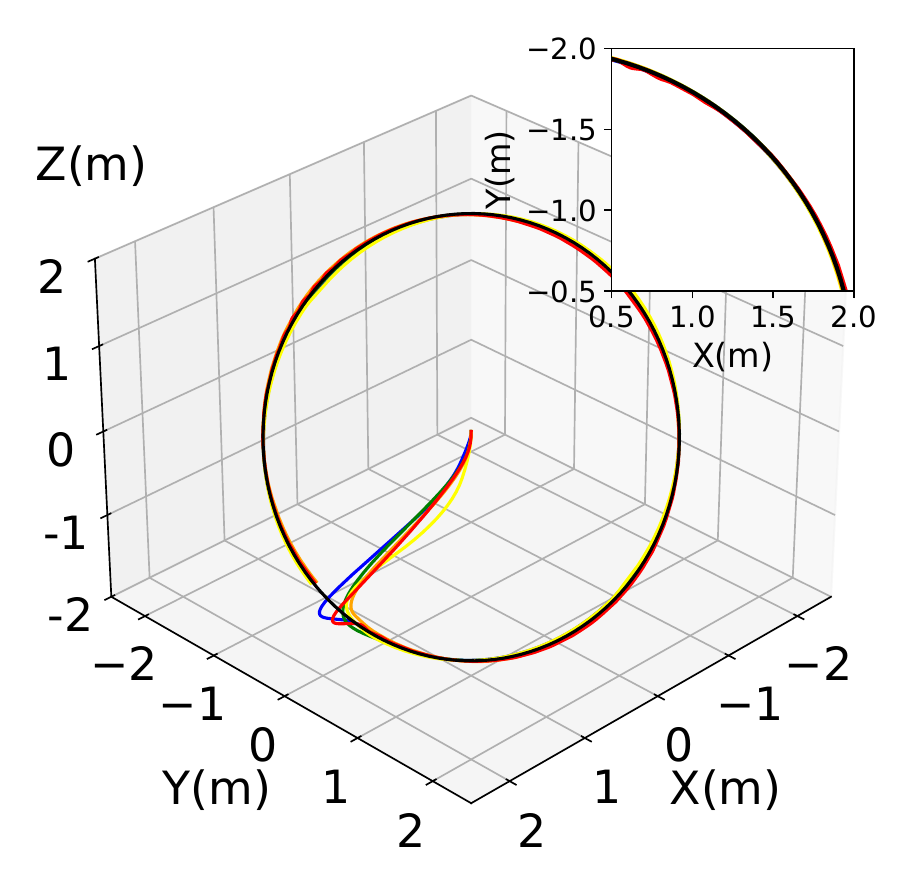}\label{fig:circle_no}}
\hfill
\subfloat{\includegraphics[width=0.22\textwidth]{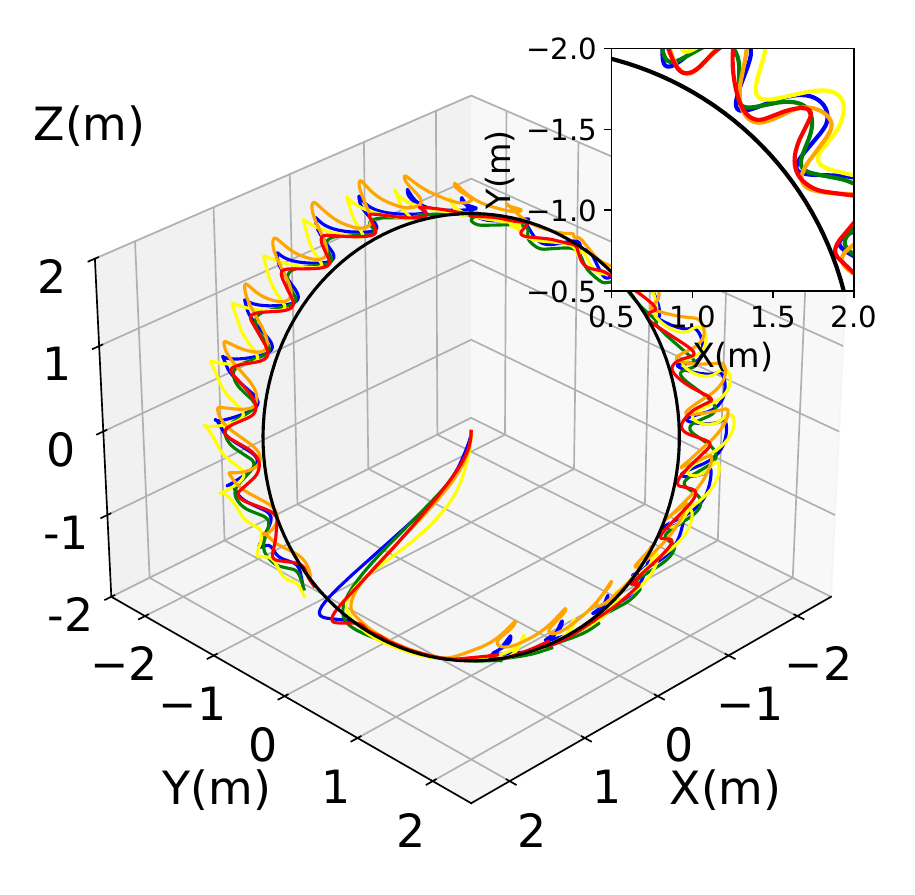}\label{fig:circle_d1}}
\hfill
\subfloat{\includegraphics[width=0.22\textwidth]{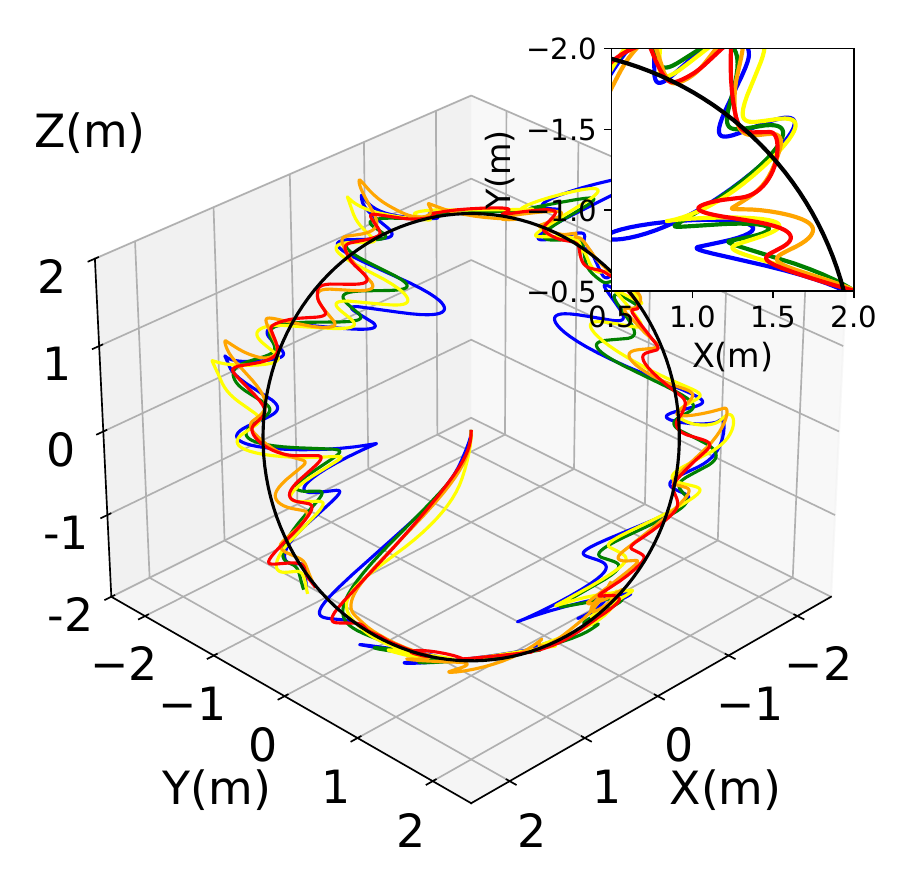}\label{fig:circle_d2}}
\hfill
\subfloat{\includegraphics[width=0.22\textwidth]{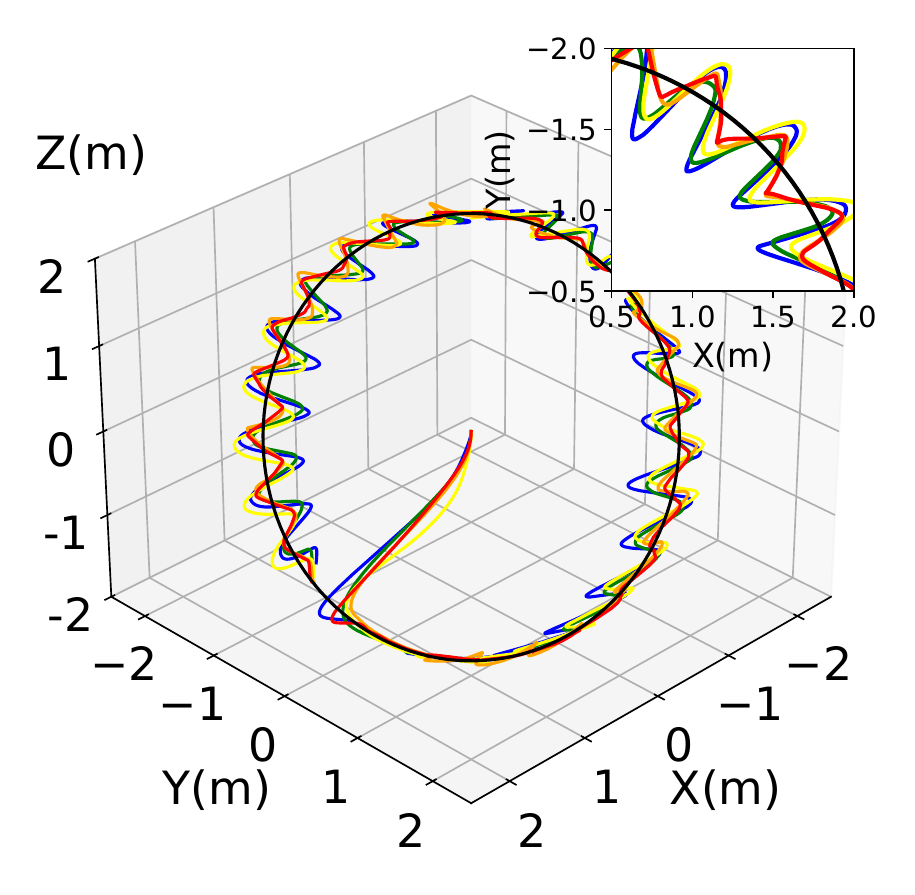}\label{fig:circle_d3}}
\vspace{0.1mm}
\includegraphics[width=0.9\textwidth]{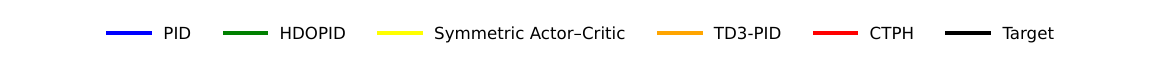}
\vspace{-5mm}
\caption{The results of the controller performing elliptical trajectory tracking under different environmental conditions. From left to right, the scenarios correspond to: no disturbance, disturbance \(d_1\), disturbance \(d_2\), and disturbance \(d_3\).}
\label{fig:all_pictures_circle}
\end{figure*}

\begin{figure*}[!t]
\centering
\subfloat{\includegraphics[width=0.22\textwidth]{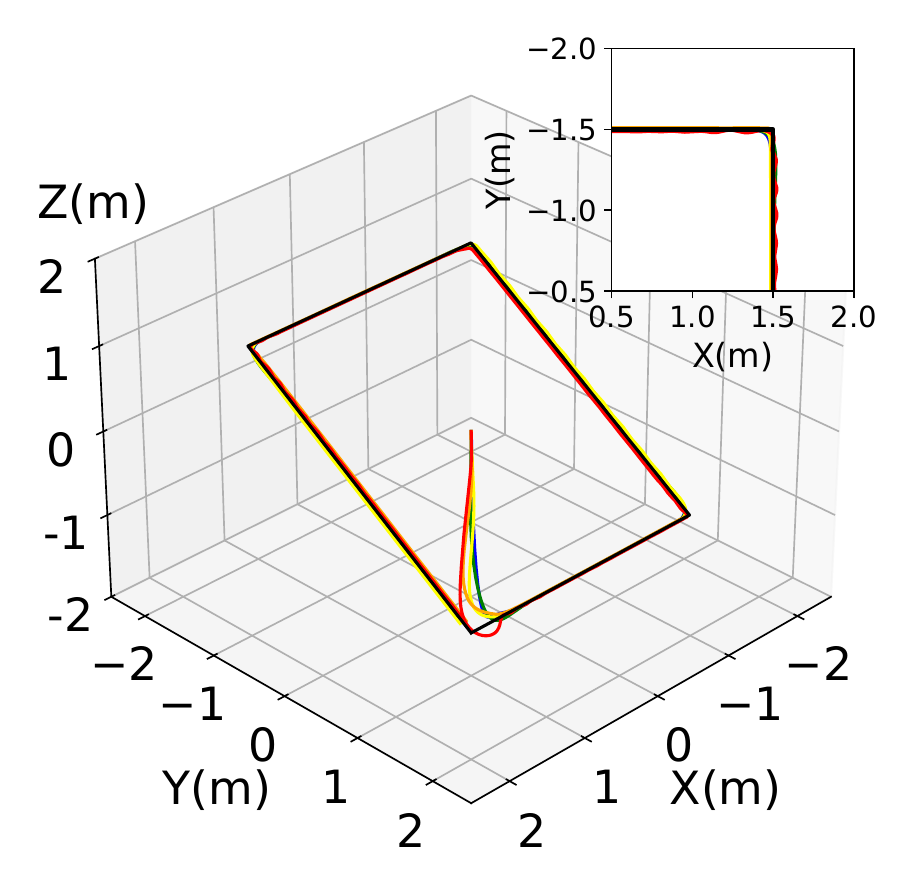}\label{fig:rectangle_no}}
\hfill
\subfloat{\includegraphics[width=0.22\textwidth]{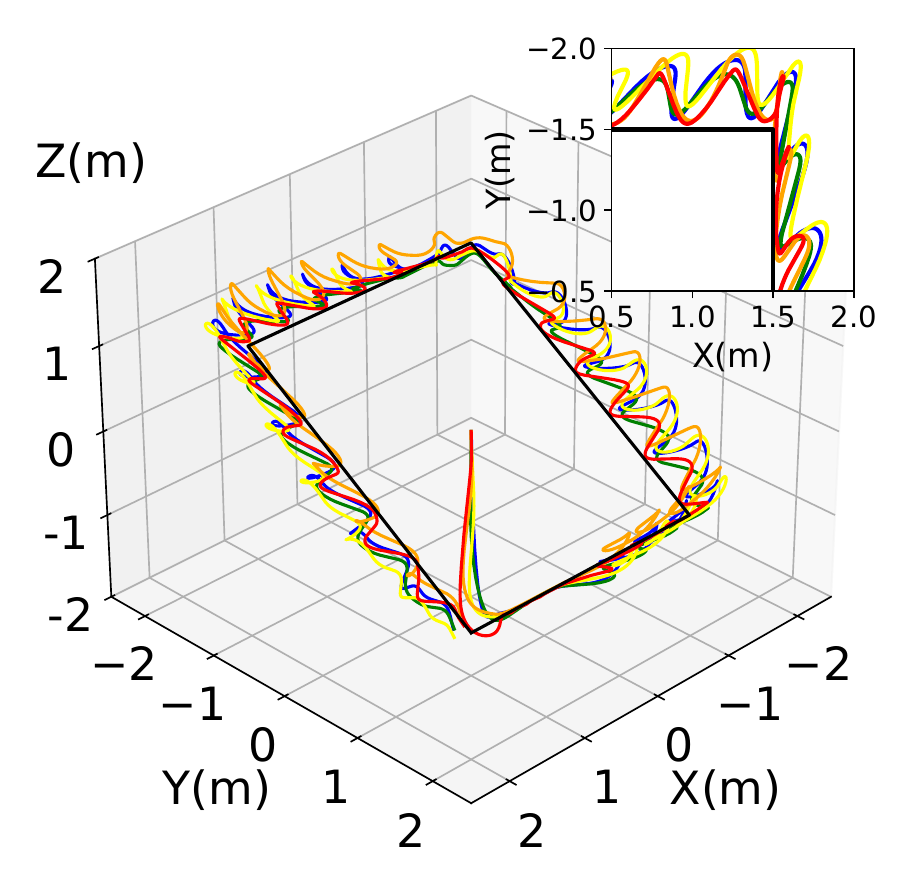}\label{fig:rectangle_d1}}
\hfill
\subfloat{\includegraphics[width=0.22\textwidth]{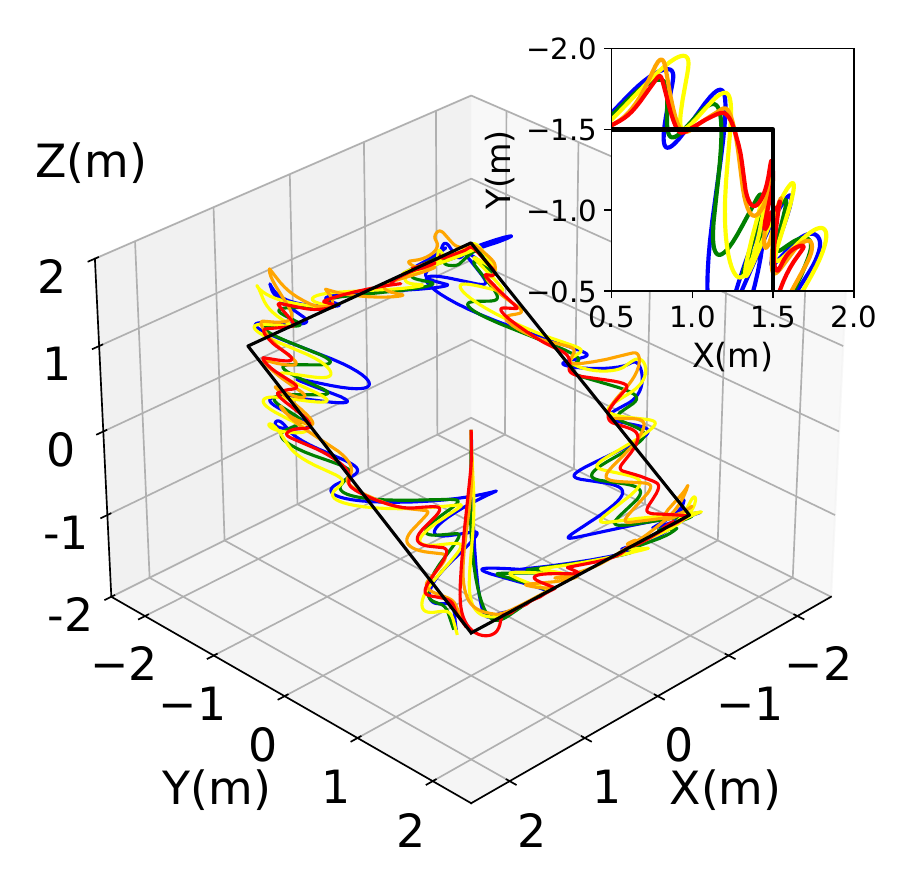}\label{fig:rectangle_d2}}
\hfill
\subfloat{\includegraphics[width=0.22\textwidth]{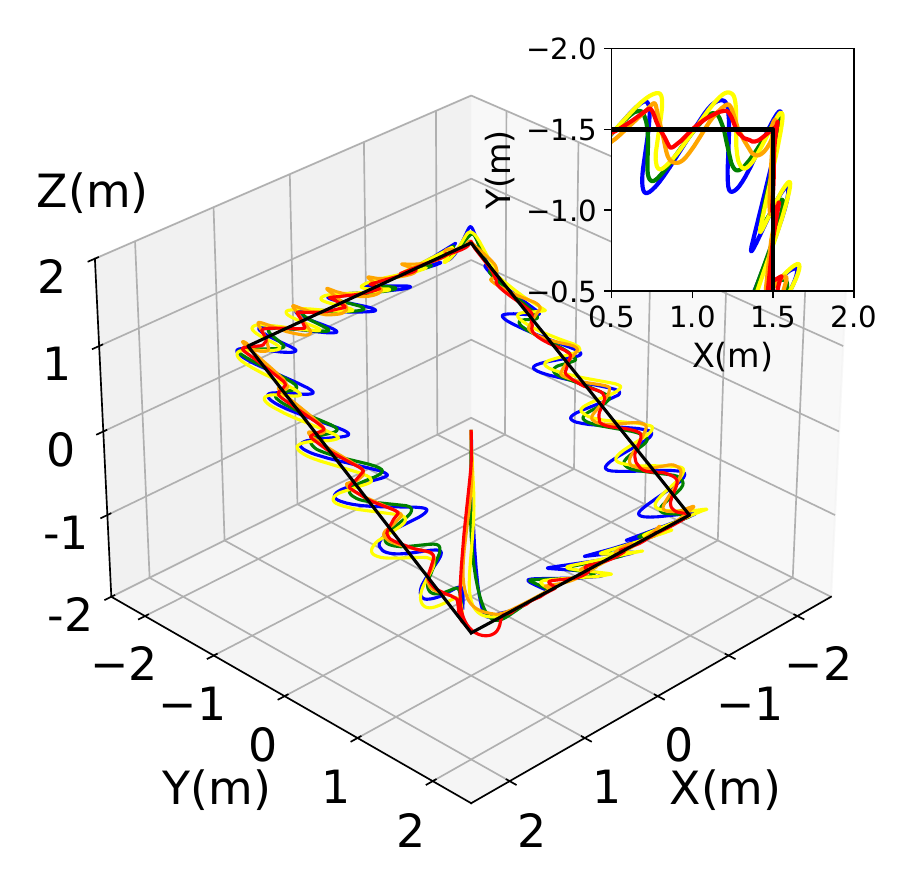}\label{fig:rectangle_d3}}
\vspace{0.1mm}
\includegraphics[width=0.9\textwidth]{figure_legend.pdf}
\vspace{-5mm}
\caption{The results of the controller performing rectangular trajectory tracking under different environmental conditions. From left to right, the scenarios correspond to: no disturbance, disturbance \(d_1\), disturbance \(d_2\), and disturbance \(d_3\).}
\label{fig:all_pictures_rectangle}
\end{figure*}

\subsection{Complex Time-Varying Trajectory Tracking}
\label{sec:V-E}
To evaluate tracking performance under identical conditions, two representative trajectories were used: the elliptical path in \cite{ref36} and a closed square-shaped spatial path connecting the vertices $(1.5,\, 1.5,\, -1)$, $(-1.5,\, 1.5,\, -1)$, $(-1.5,\, -1.5,\, 1)$, and $(1.5,\, -1.5,\, 1)$. 
For the latter, the yaw angle $\psi$ varied sinusoidally within $\pm 120^\circ$, increasing the control difficulty, especially at corners. To assess robustness, three representative wind disturbance models with predefined time-varying magnitudes and patterns were introduced. The disturbance models are defined as follows:
\begin{equation}
\begin{aligned}
d_{1}(t) &= 0.05 + 0.05\sin(\pi t + \frac{\pi}{3}),\\
d_{2}(t) &= 0.05\sin\Bigl(\frac{\pi}{4} t\Bigr) + 0.05\sin(\pi t + \frac{\pi}{3}),\\
d_{3}(t) &= \mathcal{N}(0, 0.03^2) + 0.05\sin(\pi t + \frac{\pi}{3}).
\end{aligned}
\label{eq:disturbances}
\end{equation}

To assess tracking performance, the position error was defined as $\bm{e}(t) = \bm{p}(t) - \bm{p}^{d}(t)$, where $\bm{p}^{d}(t)$ and $\bm{p}(t)$ denote the desired and actual trajectories, respectively.
Based on this error, three metrics were used. The Root Mean Square Error (RMSE), expressed as \(\mathrm{RMSE}_i = \sqrt{\frac{1}{N} \sum\limits_{\mathclap{t=1}}^{\mathclap{N}} e_i^2(t)}, \quad i \in \{x, y, z\}\), measures the overall error magnitude and emphasizes large deviations, 
whereas the Mean Absolute Error (MAE), given by \(\mathrm{MAE}_i = \frac{1}{N} \sum\limits_{\mathclap{t=1}}^{\mathclap{N}} |e_i(t)|\), measures the average error magnitude and is less sensitive to outliers. Temporal performance was further evaluated by the trajectory tracking latency. 
Following the metric in~\cite{ref37}, the latency at each time step was estimated by matching the current actual position to the closest reference point within a finite historical window:
\begin{equation}
L(T) = \Bigl(
T - \mathop{\arg\min}\limits_{K \in [T-W,\,T)} 
\left\| \bm{p}(T) - \bm{p}^{d}(K) \right\|_2
\Bigr) \Delta t,
\end{equation}
where $T$ is the current discrete time step, $\Delta t$ is the sampling period, and $W$ is the historical search-window length. Here, $\bm{p}(T)$ and $\bm{p}^{d}(K)$ denote the actual and reference positions, respectively.
The minimizing index corresponds to the closest reference point within the window and is used to quantify the tracking latency. For overall evaluation, the mean latency over the trajectory, denoted by $L_{\mathrm{avg}}$, is obtained by averaging $L(T)$ over all time steps.
The proposed CTPH controller was compared with four alternative controllers within a cascaded control architecture:
\begin{enumerate}
    \item PID: a cascaded PID controller with tuned gains.
    \item HDOPID: a cascaded PID controller with the hybrid disturbance observer.
    \item Symmetric AC \cite{ref36}: a cascaded controller based on a symmetric actor-critic reinforcement learning framework.
    \item TD3-PID: a TD3-PID hybrid controller where the improved TD3 regulates horizontal position and baseline PID regulates attitude and altitude.
    \item CTPH: the proposed method.
\end{enumerate}

\begin{table*}[!t]
\caption{Tracking Performance of the Controllers Under Complex Trajectories. Values to the Left of the Slash Correspond to Elliptical Trajectories, While Those to the Right Correspond to Rectangular Trajectories.}
\label{tab:tracking_performance} 
\scriptsize
\centering
\begin{tabular*}{\textwidth}{@{\extracolsep{\fill}} c c c c c c c c c}
\toprule
\multicolumn{2}{c}{\textbf{Variable}} & \multicolumn{3}{c}{\textbf{RMSE (m)}} & \multicolumn{3}{c}{\textbf{MAE (m)}} & $\bm{L_{\mathrm{avg}}}\ \textbf{(s)}$ \\
\cmidrule(lr){3-5} \cmidrule(lr){6-8}
 & & x & y & z & x & y & z &  \\
\midrule
\multirow{4}{*}{PID}
 & No & 0.140\slsh 0.111 & 0.129\slsh 0.141 & 0.070\slsh 0.063 & 0.128\slsh 0.071 & 0.114\slsh 0.112 & 0.065\slsh 0.050 & 0.72\slsh 0.71 \\
 & $d_{1}(t)$ & 0.168\slsh 0.203 & 0.285\slsh 0.319 & 0.184\slsh 0.186 & 0.133\slsh 0.163 & 0.237\slsh 0.272 & 0.151\slsh 0.155 & 0.48\slsh 0.56 \\
 & $d_{2}(t)$ & 0.483\slsh 0.443 & 0.354\slsh 0.332 & 0.109\slsh 0.109 & 0.323\slsh 0.289 & 0.256\slsh 0.236 & 0.085\slsh 0.083 & 0.68\slsh 0.56 \\
 & $d_{3}(t)$ & 0.258\slsh 0.232 & 0.195\slsh 0.190 & 0.070\slsh 0.064 & 0.207\slsh 0.180 & 0.160\slsh 0.155 & 0.065\slsh 0.052 & 0.69\slsh 0.61 \\
\midrule
\multirow{4}{*}{HDOPID} 
 & No & 0.140\slsh 0.112 & 0.129\slsh 0.143 & 0.093\slsh 0.084 & 0.128\slsh 0.072 & 0.114\slsh 0.113 & 0.087\slsh 0.067 & 0.76\slsh 0.75 \\
 & $d_{1}(t)$ & 0.125\slsh 0.161 & 0.253\slsh 0.285 & \textbf{0.093\slsh 0.085} & 0.102\slsh 0.132 & 0.215\slsh 0.247 & 0.086\slsh \textbf{0.069} & 0.49\slsh 0.60 \\
 & $d_{2}(t)$ & 0.342\slsh 0.303 & 0.251\slsh 0.233 & 0.093\slsh 0.084 & 0.244\slsh 0.214 & 0.201\slsh 0.185 & 0.087\slsh 0.068 & 0.71\slsh 0.60 \\
 & $d_{3}(t)$ & 0.214\slsh 0.187 & 0.160\slsh 0.160 & 0.093\slsh 0.084 & 0.181\slsh 0.148 & 0.131\slsh 0.131 & 0.087\slsh 0.068 & 0.75\slsh 0.70 \\
\midrule
\multirow{4}{*}{Symmetric AC} 
 & No & 0.108\slsh 0.085 & 0.099\slsh 0.110 & \textbf{0.025\slsh 0.023} & 0.097\slsh 0.058 & 0.086\slsh 0.087 & \textbf{0.023\slsh 0.018} & 0.51\slsh 0.51 \\
 & $d_{1}(t)$ & 0.213\slsh 0.247 & 0.311\slsh 0.344 & 0.158\slsh 0.158 & 0.166\slsh 0.201 & 0.261\slsh 0.294 & 0.121\slsh 0.121 & \textbf{0.35\slsh 0.45} \\
 & $d_{2}(t)$ & 0.344\slsh 0.316 & 0.271\slsh 0.263 & \textbf{0.082\slsh 0.083} & 0.241\slsh 0.221 & 0.210\slsh 0.200 & \textbf{0.049\slsh 0.048} & 0.56\slsh 0.47 \\
 & $d_{3}(t)$ & 0.182\slsh 0.166 & 0.171\slsh 0.171 & \textbf{0.030\slsh 0.028} & 0.150\slsh 0.134 & 0.141\slsh 0.139 & \textbf{0.025\slsh 0.024} & 0.53\slsh 0.51 \\
\midrule
\multirow{4}{*}{TD3-PID} 
 & No & 0.066\slsh 0.049 & 0.066\slsh 0.071 & 0.087\slsh 0.079 & 0.058\slsh \textbf{0.034} & 0.057\slsh 0.058 & 0.081\slsh 0.063 & 0.41\slsh 0.41 \\
 & $d_{1}(t)$ & 0.160\slsh 0.188 & 0.263\slsh 0.299 & 0.292\slsh 0.293 & 0.116\slsh 0.139 & 0.204\slsh 0.239 & 0.257\slsh 0.262 & 0.39\slsh 0.48 \\
 & $d_{2}(t)$ & 0.207\slsh 0.181 & 0.183\slsh 0.177 & 0.164\slsh 0.166 & 0.144\slsh 0.126 & 0.138\slsh 0.129 & 0.126\slsh 0.126 & \textbf{0.49\slsh 0.45} \\
 & $d_{3}(t)$ & 0.121\slsh 0.108 & 0.121\slsh 0.127 & 0.099\slsh 0.093 & 0.096\slsh 0.083 & 0.095\slsh 0.098 & 0.083\slsh 0.077 & 0.48\slsh 0.48 \\
\midrule
\multirow{4}{*}{CTPH} 
 & No & \textbf{0.061\slsh 0.048} & \textbf{0.060\slsh 0.065} & 0.087\slsh 0.079 & \textbf{0.054}\slsh 0.035 & \textbf{0.053\slsh 0.055} & 0.081\slsh 0.063 & \textbf{0.39\slsh 0.40} \\
 & $d_{1}(t)$ & \textbf{0.099\slsh 0.133} & \textbf{0.210\slsh 0.258} & 0.096\slsh 0.091 & \textbf{0.072\slsh 0.097} & \textbf{0.161\slsh 0.201} & \textbf{0.081}\slsh 0.078 & 0.36\slsh 0.47 \\
 & $d_{2}(t)$ & \textbf{0.196\slsh 0.173} & \textbf{0.133\slsh 0.147} & 0.087\slsh 0.083 & \textbf{0.135\slsh 0.114} & \textbf{0.101\slsh 0.107} & 0.081\slsh 0.071 & 0.50\slsh \textbf{0.45} \\
 & $d_{3}(t)$ & \textbf{0.102\slsh 0.096} & \textbf{0.090\slsh 0.105} & 0.087\slsh 0.080 & \textbf{0.085\slsh 0.072} & \textbf{0.073\slsh 0.082} & 0.080\slsh 0.066 & \textbf{0.45\slsh 0.47} \\
\bottomrule
\end{tabular*}
\end{table*}

\begin{figure*}[!t]
\centering
\subfloat[]{\includegraphics[width=0.32\textwidth]{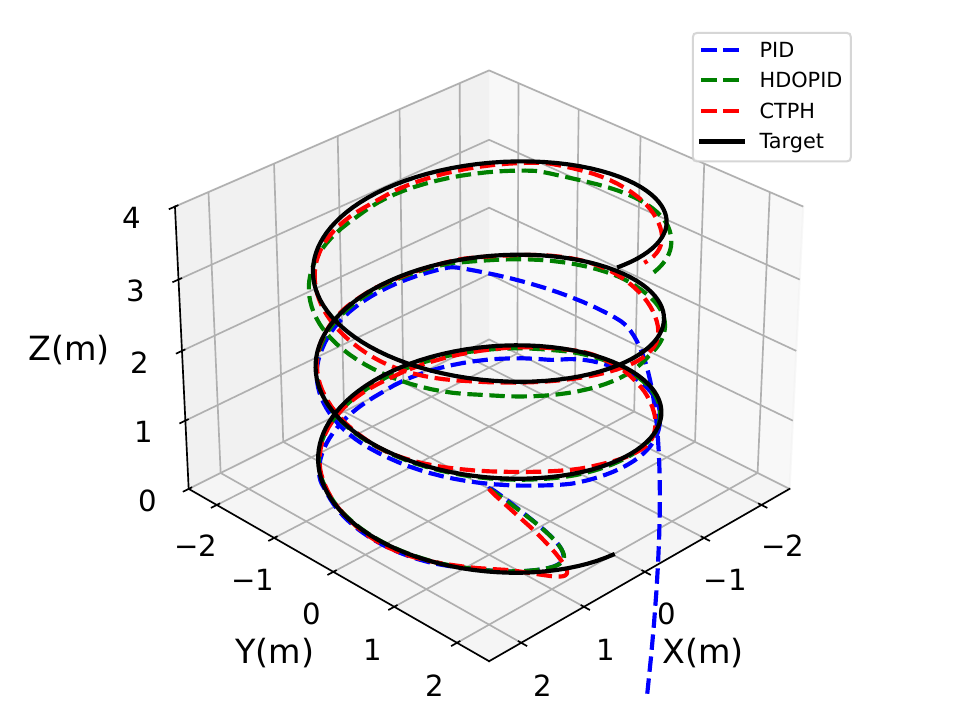}\label{fig:pid}}
\subfloat[]{\includegraphics[width=0.67\textwidth]{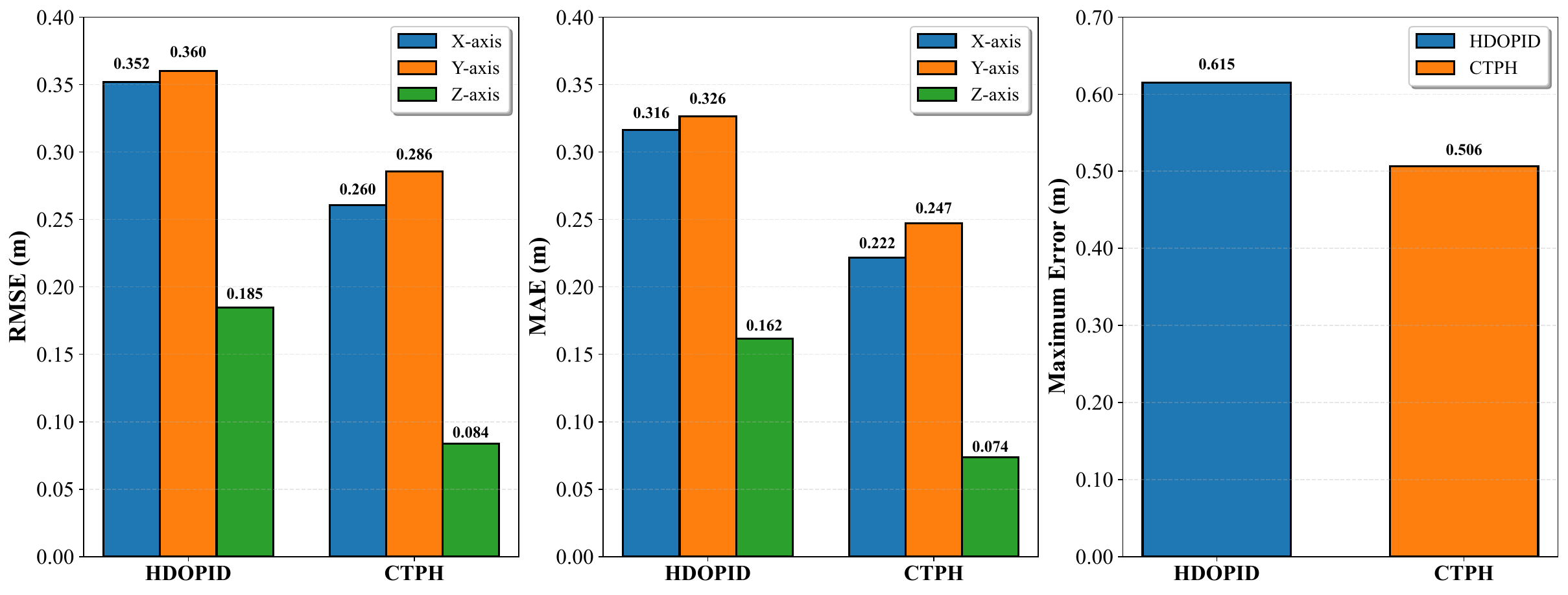}\label{fig:rmse}}
\caption{The generalization capability of the proposed strategy was evaluated. (a) shows the trajectory tracking performance of the PID, HDOPID, and CTPH controllers under variations in physical parameters, while (b) compares the RMSE, MAE, and maximum error between HDOPID and CTPH.}
\label{fig:all_pictures_change}
\end{figure*}

In the trajectory testing environment, the performance of the five controllers under different wind disturbance conditions is illustrated in Fig.~\ref{fig:all_pictures_circle} and Fig.~\ref{fig:all_pictures_rectangle}. 
As summarized in Table~\ref{tab:tracking_performance}, the proposed CTPH method generally achieves lower RMSE, MAE, and average latency values than the other controllers, indicating accurate trajectory tracking under complex disturbances. 
Although the Symmetric AC controller shows competitive performance in some cases, particularly in suppressing z-axis deviations, and TD3-PID improves upon PID and HDOPID in several scenarios, both methods are generally more sensitive to time-varying disturbances. 
Overall, these results highlight the advantage of CTPH as a disturbance-aware control strategy that provides consistently high tracking accuracy and robustness.

Additionally, the generalization capability of the control strategy was evaluated on a quadrotor model subjected to random variations in mass and moment of inertia during a continuous trajectory-tracking task. 
To simulate structural disturbances and uncertainties, the mass and inertia parameters of the quadrotor were randomly rescaled every $3$ seconds. The results of CTPH, PID, and HDOPID are shown in Fig.~\ref{fig:all_pictures_change}. 
As shown in Fig.~\ref{fig:all_pictures_change}\subref{fig:pid}, when the mass and moment of inertia vary randomly within a 30\% range, the PID controller fails to complete the trajectory-tracking task, whereas HDOPID and CTPH remain relatively stable. 
Fig.~\ref{fig:all_pictures_change}\subref{fig:rmse} further shows that CTPH achieves smaller tracking errors than HDOPID, indicating better generalization performance.

\subsection{Real-World Flight Experiments}
\label{sec:V-F}

\begin{figure}[!t]
    \centering
    \includegraphics[width=0.45\textwidth]{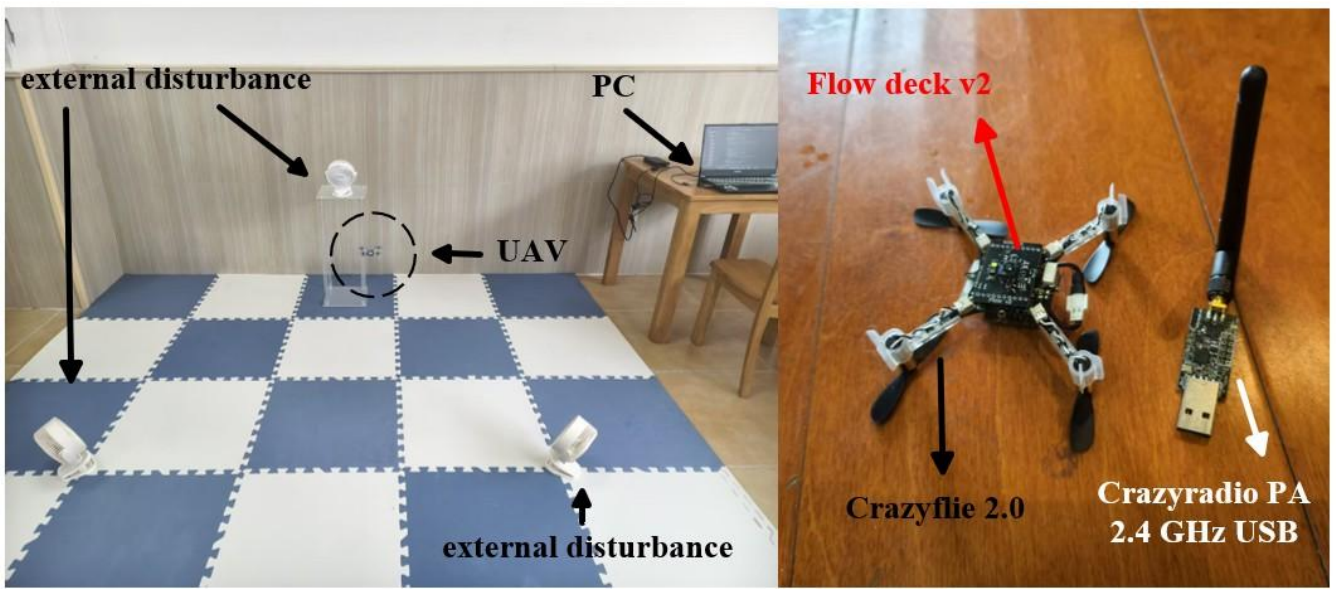}
    \caption{Introduction to the experimental platform and deployment of experimental scenarios.}
    \label{fig:environment}
\end{figure}

\begin{table*}[!t]
\captionsetup{justification=raggedright,singlelinecheck=false}
\caption{Comparison of Performance Metrics Between the PID Controller and the CTPH Controller Under Different Test Conditions}
\label{tab:datas} 
\scriptsize
\setlength{\tabcolsep}{4pt}
\begin{tabular*}{\textwidth}{@{\extracolsep{\fill}} l l c c c c c c c c @{}}
    \toprule
    \multirow{2}{*}{\textbf{Controller}} 
    & \multirow{2}{*}{\textbf{Axis}}
    & \multicolumn{4}{c}{\textbf{Stable environment}} 
    & \multicolumn{4}{c}{\textbf{Wind-disturbance environment}} \\ 
    \cmidrule(lr){3-6} \cmidrule(lr){7-10}
     & & \textbf{RMSE (m)} & \textbf{MAE (m)} & \textbf{MaxErr (m)} & \textbf{MinErr (m)} 
     & \textbf{RMSE (m)} & \textbf{MAE (m)} & \textbf{MaxErr (m)} & \textbf{MinErr (m)} \\ 
    \midrule
    \multirow{3}{*}{PID}
    & x & 0.074 & 0.061 & \textbf{0.339} & 4.0e-5 & 0.081 & 0.063 & \textbf{0.262} & 5.0e-5 \\ 
    & y & 0.052 & 0.044 & \textbf{0.120} & 3.0e-5 & 0.092 & 0.074 & 0.254 & 3.0e-5 \\ 
    & z & 0.079 & 0.026 & 0.505 & \textbf{0} & 0.093 & 0.049 & 0.576 & 5.0e-5 \\ 
    \midrule
    \multirow{3}{*}{CTPH}
    & x & \textbf{0.051} & \textbf{0.029} & 0.347 & \textbf{2.0e-5} & \textbf{0.052} & \textbf{0.034} & 0.272 & \textbf{0} \\ 
    & y & \textbf{0.033} & \textbf{0.027} & 0.133 & \textbf{2.0e-5} & \textbf{0.049} & \textbf{0.036} & \textbf{0.208} & \textbf{1.0e-5} \\ 
    & z & \textbf{0.076} & \textbf{0.024} & \textbf{0.502} & \textbf{0} & \textbf{0.078} & \textbf{0.025} & \textbf{0.509} & \textbf{0} \\ 
    \bottomrule
\end{tabular*}
\end{table*}

\begin{figure}[!t]
    \centering
    \includegraphics[width=0.4\textwidth]{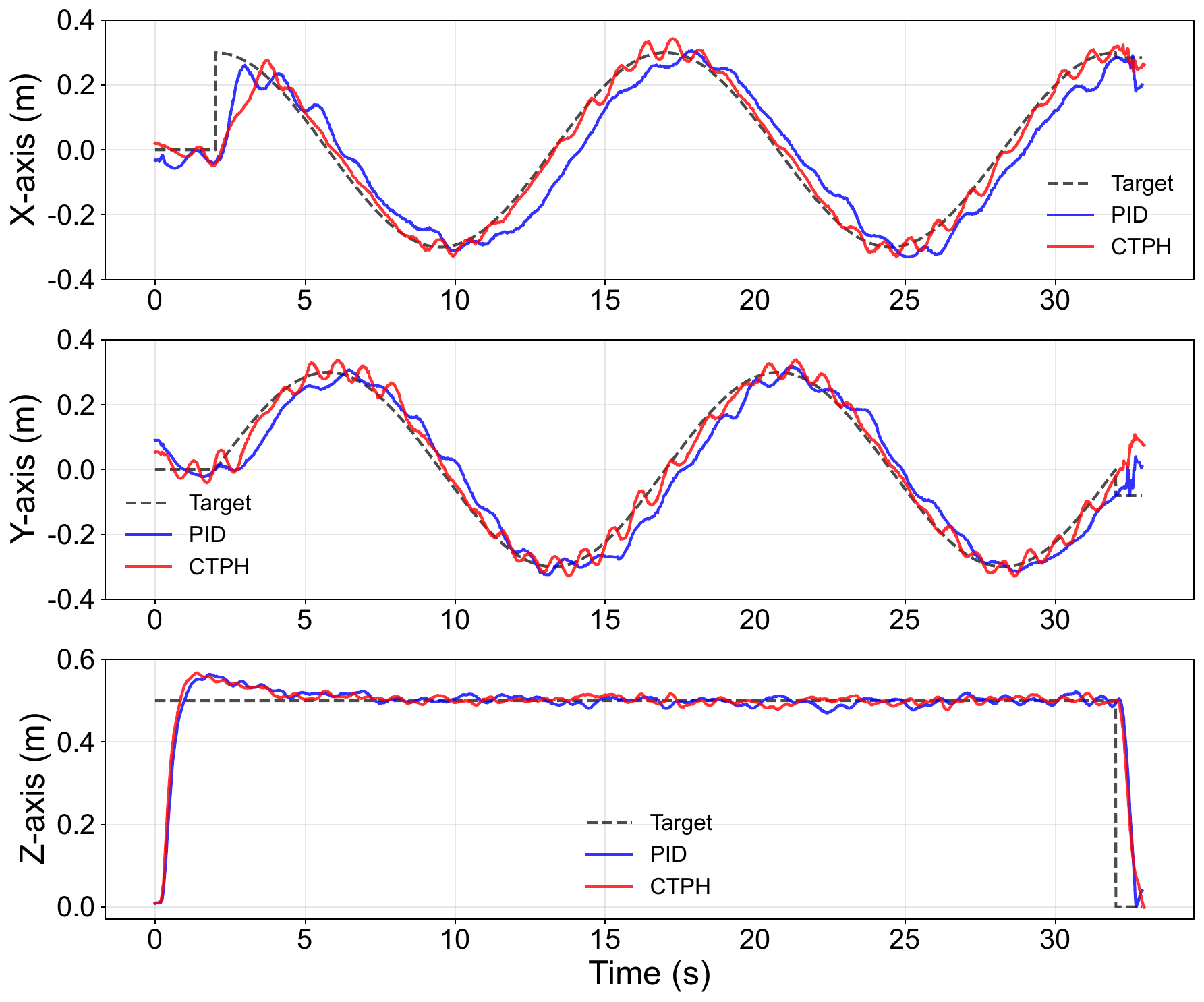}
    \caption{Trajectory tracking comparison between the PID controller and the CTPH controller along the X, Y, and Z axes under a stable environment.}
    \label{fig:cfflyer_test}
\end{figure}

\begin{figure}[!t]
    \centering
    \includegraphics[width=0.4\textwidth]{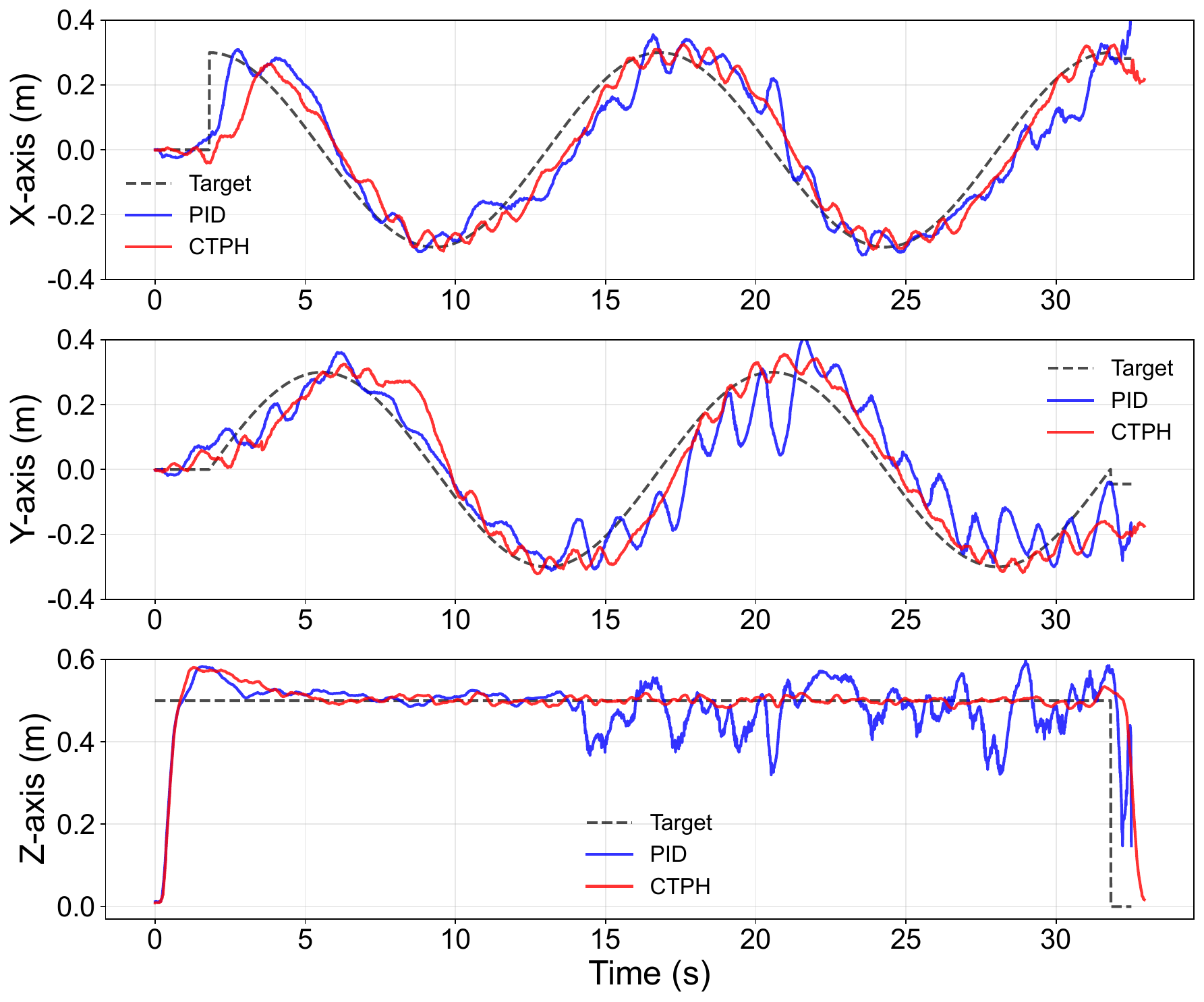}
    \caption{Trajectory tracking comparison between the PID controller and the CTPH controller along the X, Y, and Z axes under a wind-disturbance environment.}
    \label{fig:cfflyer_wind_test}
\end{figure}

In this section, real-world trajectory tracking experiments were conducted to verify the effectiveness of the proposed method. The experimental platform was a Crazyflie 2.0 quadrotor equipped with an onboard IMU and a Flow Deck v2 for optical-flow-based velocity estimation and height sensing. 
The system was controlled by an onboard STM32 microprocessor and communicated with the host computer through a Crazyradio PA module. All experiments were conducted indoors, and only onboard sensors were used for state estimation without an external tracking system. 
To improve estimation reliability, the UAV position and velocity were estimated by fusing IMU and Flow Deck v2 measurements with an extended Kalman filter (EKF). The experimental platform and test environment are shown in Fig.~\ref{fig:environment}.

To reduce the discrepancy between simulation and real-world experiments, the simulation model was refined according to the actual UAV dynamics and control architecture. In the Crazyflie firmware, the desired horizontal acceleration generated by the TD3 controller in simulation was reinterpreted as a desired horizontal velocity command on the real platform. 
In addition, the control frequencies of the position and attitude loops were synchronized with the real control cycle, and communication delay as well as sensor noise were incorporated to improve the robustness of Sim2Real transfer. The designed disturbance observer was implemented in the attitude control layer of the Crazyflie firmware.

The proposed strategy was evaluated on the real Crazyflie platform in a horizontal circular trajectory-tracking task at a fixed altitude of 0.5\,m. The reference trajectory was centered at $(0,0)$ with a radius of 0.3\,m, completed in 15\,s per revolution, and repeated for two full laps. 
Two scenarios were considered: a calm indoor environment and an environment with artificially induced wind disturbances. A PID controller was used as the baseline.
Fig.~\ref{fig:cfflyer_test} and Fig.~\ref{fig:cfflyer_wind_test} show the trajectory-tracking performance of the PID and CTPH controllers under the two test conditions, while Table~\ref{tab:datas} reports the corresponding RMSE, MAE, MaxErr, and MinErr values. 
Under calm conditions, CTPH achieved better overall tracking performance, especially in the horizontal plane, than the conventional PID controller. Under wind disturbances, CTPH still maintained satisfactory tracking performance, whereas larger deviations were observed with PID at some time instants. These results support the effectiveness of the proposed controller for real-world trajectory tracking under external disturbances. Nevertheless, weaker tracking performance was observed in the initial stage of the trajectory, which contributed to relatively lower tracking accuracy along the \(x\)-axis than along the \(y\)-axis.

\section{Conclusion}
\label{sec:VI}
In this article, a cascaded hybrid control framework combining TD3-based reinforcement learning with PID control is developed for quadrotor trajectory tracking. 
Key enhancements include an aggregated Q-network, a PID-guided expert policy, and a reward-filtered experience replay to improve training stability and sample efficiency. 
A structurally simple hybrid disturbance observer is integrated into the PID layer to strengthen disturbance rejection and ensure robust attitude and altitude control. 
Simulation results show that the proposed method achieves faster convergence, smaller tracking errors, and improved robustness compared with baseline TD3 and PID controllers. 
Real-world experiments further confirm its ability to maintain precise and stable trajectory tracking under various conditions. A limitation is that the learning process still relies on trial-and-error interaction. 
Future work will explore model-aware reinforcement learning schemes and tighter integration of physical insights to enhance stability and interpretability.

\end{document}